 \documentclass[final,5p,times]{elsarticle}

\usepackage{amssymb}

\usepackage[utf8]{inputenc}
\usepackage{graphicx}
\usepackage{epsf}
\usepackage{bm}
\usepackage{amsmath}
\usepackage{amsfonts}
\usepackage{amssymb}
\usepackage{subfigure}
\usepackage{caption}
\usepackage{scalerel}
\newcommand{\abs}[1]{\left\lvert#1\right\rvert}
\usepackage[font=scriptsize,labelfont=bf]{caption}
\RequirePackage{latexsym}
\usepackage{widetext}
\usepackage{flushend}
\usepackage{cuted}
\usepackage{hyperref}
\hypersetup{
    colorlinks=true,
    linkcolor=red,
    citecolor=blue,
    filecolor=magenta,      
    urlcolor=cyan,
    pdftitle={The dynamics of scalar-field Quintom cosmological models},
    pdfpagemode=FullScreen}

\begin{document}

\begin{frontmatter}



\title{The dynamics of scalar-field Quintom cosmological models}

%
\author{
\textsc{Jonathan Tot, Balkar Yildirim, Alan Coley} \\
Department of Mathematics and Statistics,
Dalhousie University, \\ Halifax, Nova Scotia, Canada  B3H 3J5 \\
\normalsize \href{mailto:jonathan.tot@dal.ca}{jonathan.tot@dal.ca}\\
\href{mailto:bl541771@dal.ca}{bl541771@dal.ca}\\
\href{mailto:aac@mathstat.dal.ca}{aac@mathstat.dal.ca} \\
\textsc{Genly Leon}
\\ Departamento de Matem\'aticas, Universidad Cat\'olica del Norte, \\ Avda. Angamos 0610, Casilla 1280 Antofagasta, Chile \\ 
\& Institute of Systems Science, Durban University of Technology, PO 
Box 1334,
Durban 4000, South Africa\\ 
\normalsize \href{mailto:genly.leon@ucn.cl}{genly.leon@ucn.cl}
}

\begin{abstract}
We shall present a complete (compactified) dynamical systems analysis of the Quintom model comprised of an interacting quintessence scalar field and a phantom. We find a range for the model parameters $\kappa, \lambda$ such that
there are expanding Quintom cosmologies that undergo two inflationary periods, and this behaviour is not destabilized by spatial curvature. We also discuss a class of bouncing cosmologies. Finally, the linear cosmological perturbations are studied.
\end{abstract}


\end{frontmatter}


\section{Introduction}

Observations suggest that the universe has experienced two separate acceleration phases: an accelerated phase of expansion during its very early stage of evolution, known as inflation, which occurred before the radiation-dominated era, and a recent accelerated phase of expansion, known as late-time cosmic acceleration.
Within theoretical cosmology, scalar fields play a significant role in gravitational physics. As inflation is characterized by the scalar field responsible for the universe's early time acceleration phase, a similar model can provide dynamical terms in the gravitational field equations with anti-gravitating behaviour and can also be used as a model for the description of the acceleration of the late universe. 

We are interested here in Quintom models \cite{Elizalde:2004mq,quintom1,quintom2,quintom3,quintom4,quintom5,quintom6}, comprising two interacting scalar fields (quintessence and phantom). 
In \cite{quintom1}, a class of cosmologies with Quintom dark energy with an exponential potential and dark matter was investigated from a dynamical systems perspective, and the novel behaviour of Quintom models admitting either tracking attractors or phantom attractors was observed. This analysis was extended to arbitrary potentials \cite{quintom2}, with a particular focus on the tracking phases
in the asymptotic regime where both scalar fields diverge.
In \cite{quintom5}  radiation and dark matter as described by perfect fluids
was included. A complete dynamical analysis was presented in \cite{quintom6} for a class of exponential potentials. In particular, the unstable and centre manifold of the massless scalar field cosmology motivated by the numerical results given in \cite{quintom1} were constructed. The role of the curvature on the dynamics was also investigated, and several monotonic functions were defined on relevant invariant sets for the Quintom cosmology.

The interest in Quintom models is due to their ability to describe phantom and quintessence regimes in cosmological evolution. A related model is the Chiral generalized cosmology, whose dynamics were studied in \cite{an2}. In \cite{PalLeon2021}, a detailed analysis of the dynamics of a chiral-like cosmological model with phantom terms for four different models in a spatially flat Friedmann–Lemaître–Robertson–Walker (FLRW) background space was presented. 
When the second scalar field is phantom, the parameter of the equation of state of the cosmological fluid crosses the phantom divide line twice without the appearance of ghosts. That is the quintom model we shall study here.   

There are several methods for analyzing the equations of two-field quintom models, including, for example, using the exact analytical solutions  \cite{Chervon:2013btx, Paliathanasis:2014yfa, Paliathanasis:2018vru, Chervon:2019nwq}. The verification against observational constraints \cite{Starobinsky:2001xq, Wands:2002bn, Kaiser:2012ak, Kaiser:2013sna,Planck:2018vyg} is presented elsewhere. This paper will present a complete dynamical systems analysis of the Quintom model with the action given below (\ref{sp.01}). This analysis reproduces all of the results found in \cite{PalLeon2021} but also reveals additional possible cosmological histories consistent with the model, including bouncing universes (discussed in \S\ref{bounce}). In \S\ref{sec3}, we consider FLRW cosmologies with non-zero curvature and investigate how curvature affects the local stability of inflationary epochs. Linear cosmological perturbations are analyzed in section  \S\ref{linear-perts}. The full analysis of curvature-dominated and scaling solutions will be presented in \cite{PaperII}. 
\section{The model}
\label{sec2}

We consider the two-scalar field model with action integral
\begin{equation}
	S=\int\sqrt{-g}dx^{4}\left(  R-\frac{1}{2}g^{\mu\nu}\nabla_{\mu}\phi
	\nabla_{\nu}\phi + \frac{1}{2}g^{\mu\nu}e^{\kappa\mkern1mu\phi}\nabla_{\mu}\psi
	\nabla_{\nu}\psi-V\left(  \phi\right)  \right), \label{sp.01}
\end{equation}

\noindent where the two scalar fields $\phi\left(  x^{\mu}\right)  $ and $\psi\left(x^{\nu}\right)  $ have kinetic terms which lie on a two-dimensional manifold hyperbolic space in which the motion of the scalar field occurs. In the following, we assume that $\kappa\neq0$ and the scalar field potential is that of the hyperbolic inflaton;  that is, $V\left(  \phi\right) =V_{0}e^{\lambda\phi}$. For simplicity, $\psi$ is a massless scalar field.

For the background space, we assume the FLRW geometry
\begin{equation}
	ds^2 = -dt^2 +a^2(t)\left( \frac{ dr^2}{1-Kr^2} +r^2\left( d\theta^2 +\sin^2\theta\,d\varphi^2 \right)\right),\label{sp.02}
\end{equation}
\noindent where $K$ is the spatial curvature for the three-dimensional hypersurface. For $K=0$, we have a spatially flat universe; for $K=1$, we have a closed universe; and for $K=-1$, the line element (\ref{sp.02}) describes an open universe.

\subsection{Flat $K=0$ Field Equations}

Now we study the cosmological model with action integral (\ref{sp.01}) for the line element (\ref{sp.02}) with $K=0$. The gravitational field equations are
\begin{align}
	-&3H^2 +\frac{1}{2}\dot{\phi}^2 -\frac{1}{2}e^{\kappa\mkern1mu\phi}\dot{\psi}^2
	+V\!\left( \phi \right) =0, \label{0sp.04} \\
	2\dot{H} +&3H^2 +\frac{1}{2}\dot{\phi}^2 -\frac{1}{2}e^{\kappa\mkern1mu\phi}\dot{\psi}^2
	-V\!\left( \phi \right) =0, \label{0sp.05} \\
	\ddot{\phi}+&3H\dot{\phi} +\frac{\kappa}{2}e^{\kappa\mkern1mu\phi}\dot{\psi}^2 +\partial_\phi
	V\!\left( \phi \right) =0, \label{0sp.06} \\
	\ddot{\psi} +&3H\dot{\psi} +\kappa\mkern1mu\dot{\phi}\mkern1mu\dot{\psi} =0, \label{0sp.07}
\end{align}
\noindent where $H=\frac{\dot{a}}{a}$ is the Hubble function.

\subsection{Dynamical systems formulation}
We define
\begin{equation}
    \chi^2 = \frac{1}{2} \dot{\phi}^2 + V(\phi), \quad V(\phi)= V_0 e^{\lambda \phi},
\end{equation}
\noindent and
\begin{align}
	&h^2= \frac{3 H^2}{\chi^2}, \; \eta^2= \frac{e^{\kappa \phi}\dot{\psi}^2}{2 \chi^2}, \;  \Phi^2 = \frac{\dot{\phi}^2}{2 \chi^2}, \; \Psi= \frac{V(\phi)}{\chi^2}, \label{vars}
\end{align}
\noindent which satisfy
\begin{equation}
	h^2 + \eta^2= \Phi^2+ \Psi=1.  \label{eq102}
\end{equation}
\noindent Using \eqref{eq102}, we then  have bounded variables
\begin{equation}
	0 \leq h^2, \Phi^2, \eta^2,\Psi \leq 1.  
\end{equation}
\noindent Equations \eqref{0sp.05}, \eqref{0sp.06}, \eqref{0sp.07} become
\begin{align}
	&\frac{\dot{H}}{\chi^2}=  1-h^2-\Phi^2, \label{hubble}\\
	&\frac{\ddot{\phi}}{\chi^2}= -\left(\sqrt{6}h\Phi + \kappa\left( 1-h^2 \right) +\lambda\left( 1-\Phi^2 \right)\right), \\
	&\frac{\ddot{\psi}}{\dot{\psi}\chi}= -\left( \sqrt{6}h +2\kappa\Phi \right).
\end{align}

Introducing the time derivative $d\tau=\chi d t$, and taking the derivatives of the variables with respect to the new time variable, i.e., $f^{\prime}=df/d\tau$, the field equations become
\begin{align}
	&\Phi^\prime= -\frac{ \kappa\eta^2}{\sqrt{2}} -\sqrt{3}h\Phi-\frac{ \lambda\Psi}{\sqrt{2}} -\frac{ \Phi\dot{\chi}}{\chi^2}, \\
	&h^\prime= \frac{1}{2}\sqrt{3}\left( \eta^2-h^2-\Phi^2 +\Psi\right) -\frac{ h\dot{\chi}}{\chi^2}, \\
	&\eta^\prime= -\frac{ \kappa\eta\Phi}{\sqrt{2}} -\sqrt{3}\eta h -\frac{ \eta\dot{\chi}}{\chi^2}, \\
	&\Psi^\prime= \sqrt{2}\lambda\Phi\Psi -\frac{ 2\Psi\dot{\chi}}{\chi^2}. 
\end{align}
Next, by substituting
\begin{align}
	&\frac{\dot{\chi}}{\chi^2}= -\frac{\Phi}{\sqrt{2}} \left( \kappa\left( 1-h^2 \right) +\sqrt{6}h\Phi \right), \label{chi_de}\\
	&\eta^2= 1-h^2 ,  \quad \Psi= 1-\Phi^2 ,
\end{align}
\noindent we obtain
\begin{align}
	\Phi^\prime&=-\frac{\left(1-\Phi ^2\right)}{\sqrt{2}} \left( \kappa\left(1-h^2\right) +\sqrt{6} h \Phi +\lambda\right), \label{systA}\\
	h^\prime&= \frac{\left(1-h^2\right)}{\sqrt{2}}  \left( \kappa h\Phi +\sqrt{6}\left(1-\Phi^2\right)\right), \label{systB}
\end{align}
\noindent which describes the full dynamics. Hence, we investigate the reduced dynamical system for the vector $(\Phi, h)$ given by \eqref{systA}-\eqref{systB} defined in the phase-plane $(\Phi, h) \in [-1,1]\times [-1,1]$. 

\subsection{Local analysis}

The equilibrium points are $(\Phi,h)$:
\begin{alignat*}{2}
	A&: (1,1), &&\qquad E: \left(-\frac{\lambda}{\sqrt{6}}, 1 \right),  \\
	B&: (-1,1), &&\qquad F: \left(\frac{\lambda}{\sqrt{6}},  -1 \right), \\
	C&: (1,-1), &&\qquad G: \left(-\frac{\sqrt{6}}{\sqrt{\kappa^2 + \kappa \lambda +6}}, \frac{\kappa + \lambda}{\sqrt{\kappa^2 + \kappa \lambda +6}} \right), \\
	D&: (-1,-1), &&\qquad H: \left(\frac{\sqrt{6}}{\sqrt{\kappa^2 + \kappa \lambda +6}}, -\frac{\kappa + \lambda}{\sqrt{\kappa^2 + \kappa \lambda +6}} \right), \\
	O_{\pm}&: (\pm1,0).
\end{alignat*}
Using equations \eqref{systA} and \eqref{systB}, we can find the eigenvalues of the equilibrium points and hence deduce their local stability:
\begin{alignat*}{2}
	A&: \left(-\sqrt{2} \kappa ,\sqrt{2} \lambda +2 \sqrt{3}\right), &&\qquad E: \left(\frac{\lambda ^2-6}{2 \sqrt{3}},\frac{\lambda  (\kappa +\lambda )-6}{\sqrt{3}}\right),  \\
	B&: \left(\sqrt{2} \kappa ,2 \sqrt{3}-\sqrt{2} \lambda \right), &&\qquad F: \left(-\frac{\lambda ^2-6}{2 \sqrt{3}},\frac{6-\lambda  (\kappa +\lambda
   )}{\sqrt{3}}\right), \\
	C&: \left(-\sqrt{2} \kappa ,\sqrt{2} \lambda -2 \sqrt{3}\right), &&\qquad G: \left(\delta_+, \delta_-\right),\\
	D&: \left(\sqrt{2} \kappa ,-\sqrt{2} \lambda -2 \sqrt{3}\right), &&\qquad H: \left(\Delta_+, \Delta_-\right),
	\\
	O_{\pm}&: \left(\pm\frac{\kappa }{\sqrt{2}}, \pm\sqrt{2} (\kappa +\lambda )\right).
\end{alignat*}
where
\begin{align}
\delta_\pm= \frac{\sqrt{3} \kappa \pm \sqrt{-\kappa \left(4 \left(\kappa ^2-6\right) \lambda +8
   \kappa  \lambda ^2-27 \kappa +4 \lambda ^3\right)}}{2 \sqrt{\kappa  (\kappa +\lambda
   )+6}}, \label{delta}\\
  \Delta_\pm=  \frac{-\sqrt{3} \kappa \pm \sqrt{-\kappa \left(4 \left(\kappa ^2-6\right) \lambda +8
   \kappa  \lambda ^2-27 \kappa +4 \lambda ^3\right)}}{2 \sqrt{\kappa  (\kappa +\lambda
   )+6}}.\label{Lambda}
\end{align}
\begin{figure*}[t!]
	\centering
	\subfigure[\label{fig:1} $\lambda=0.5, \kappa=1.5$]{\includegraphics[scale=0.35]{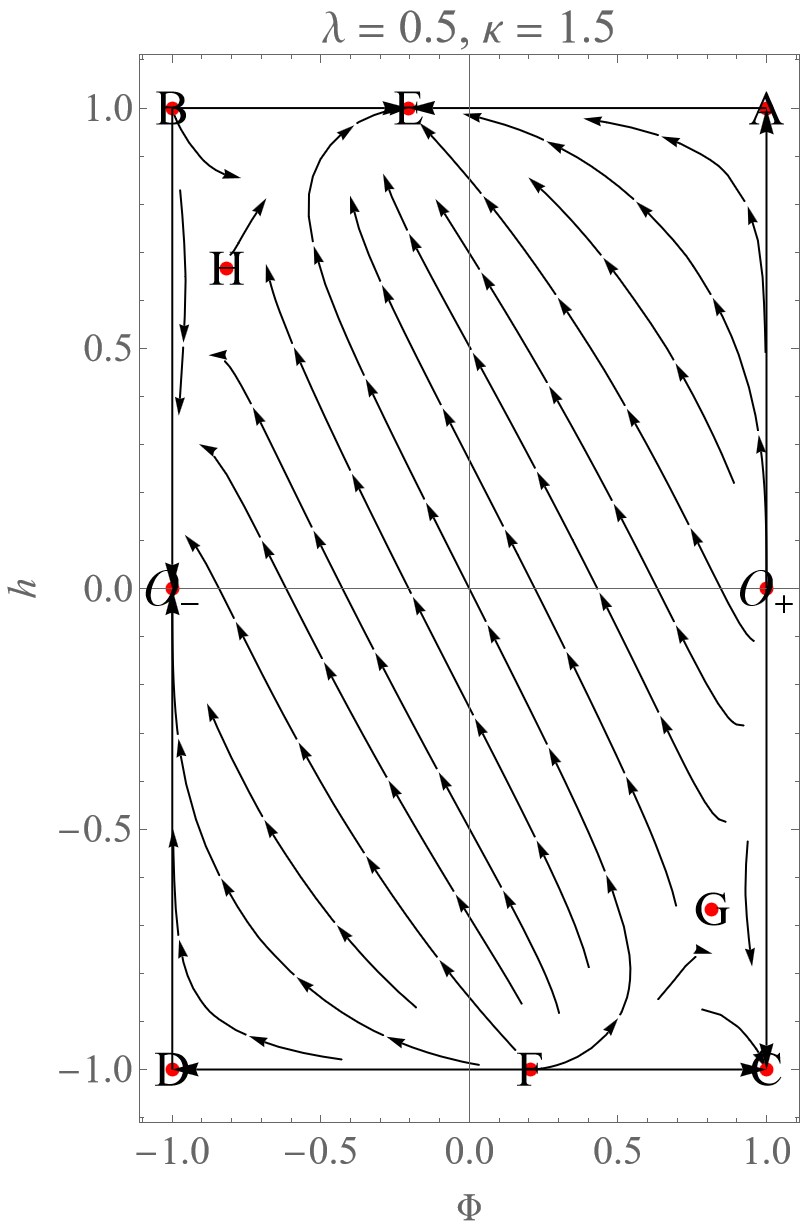}}
	\subfigure[\label{fig:1a} $\lambda=1.0, \kappa=1.5$]{\includegraphics[scale=0.35]{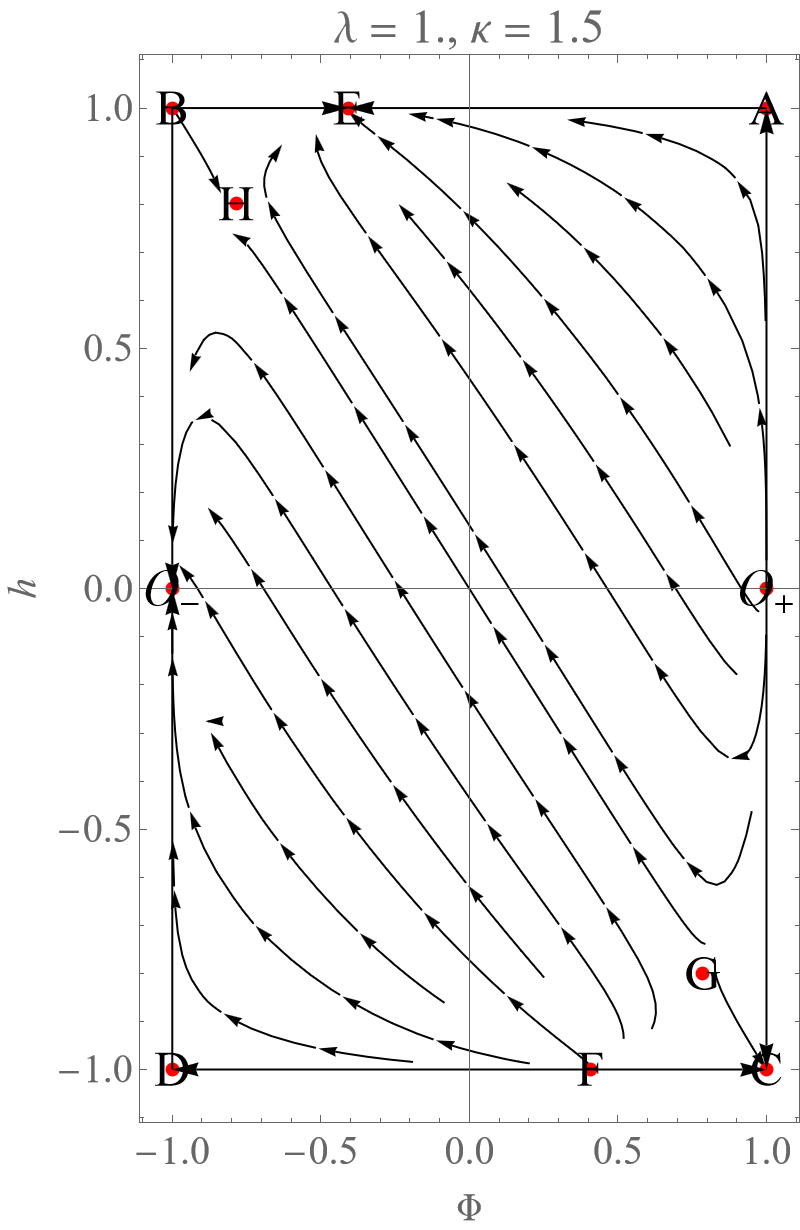}}
	\subfigure[\label{fig:1b} $\lambda=3.0, \kappa=1.5$]{\includegraphics[scale=0.35]{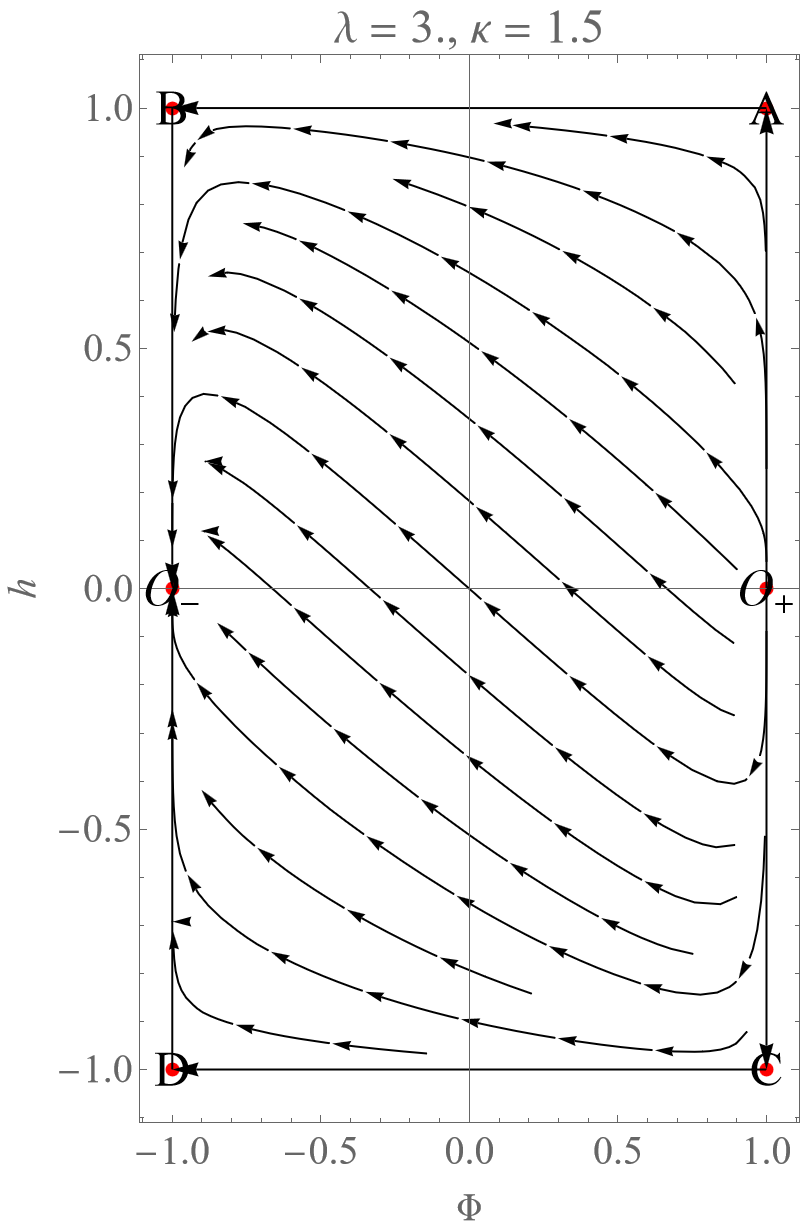}}
	\caption{Flow of the reduced  dynamical system for the vector $(\Phi, h)$ given by \eqref{systA}-\eqref{systB}  defined in the phase-plane $(\Phi, h) \in [-1,1]\times [-1,1]$.}
\end{figure*}
The results of the local stability analysis are shown in Table \ref{ep_table}. We can also determine the global dynamics and present the phase portraits. 
In Fig.~\ref{fig:1} the flow of the vector $(\Phi, h)$ given by \eqref{systA}-\eqref{systB} is presented for $\lambda=0.5, \kappa=1.5$. 
In Fig.~\ref{fig:1a},\ref{fig:1b} the parameter values used are $\lambda=1.0,\kappa=1.5$ and $\lambda=3.0, \kappa=1.5$, respectively.

\begin{figure*}[!ht]
    \centering
    \includegraphics[scale=0.55]{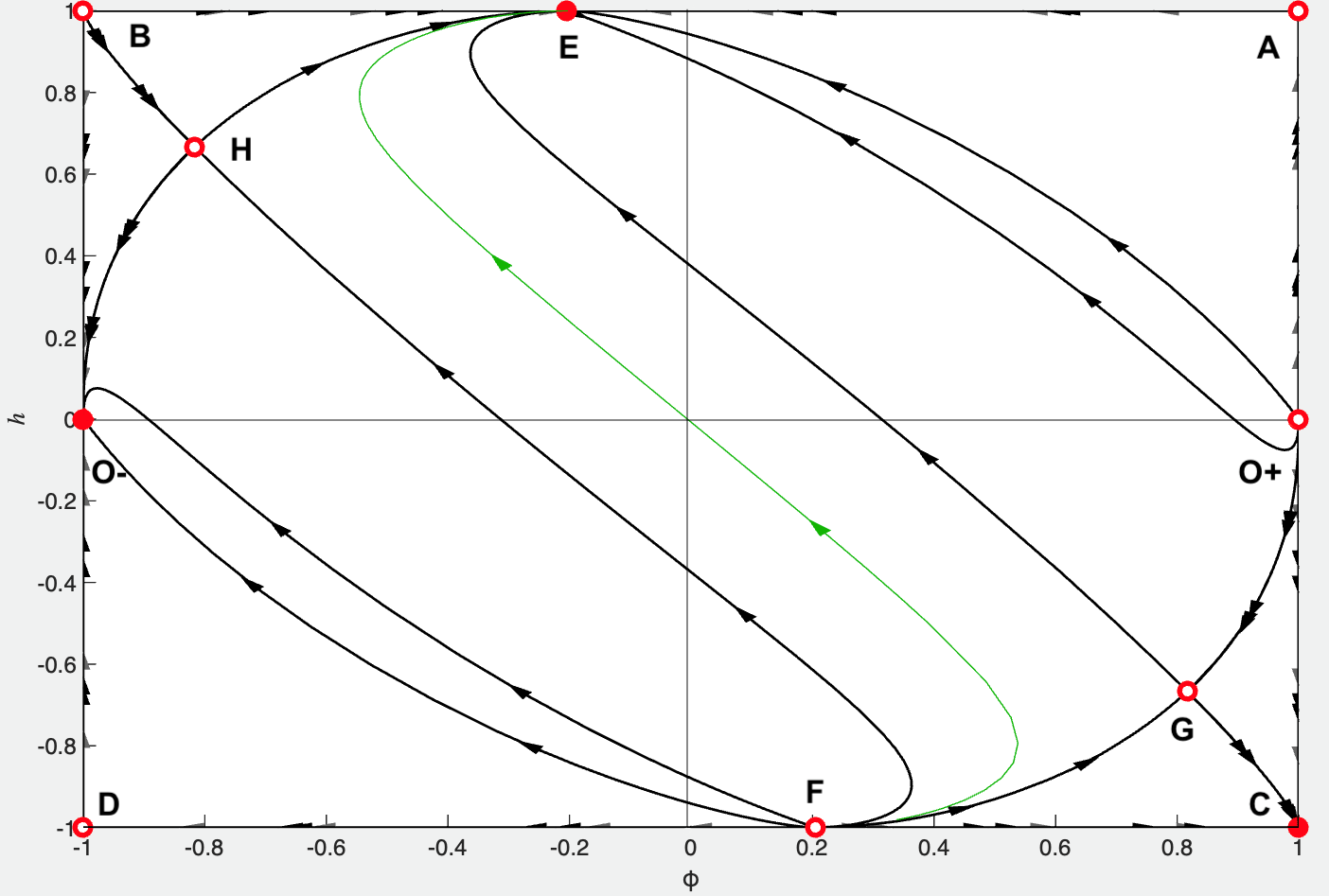}
    \caption{This figure shows a skeleton of the dynamical system \eqref{systA}-\eqref{systB} within the phase-space $[-1,1]^2$, for parameter values $\lambda=0.5,\kappa=1.5$.  The stable and unstable manifolds of the saddles $G$ and $H$ are shown, as well as orbits that are tangent to the eigenvectors of nodes (sinks $E, {O_{+}}, C$ and sources $F, {O_{-}}, B$). Shown in green is the solution which passes through the origin. This solution is completely time-symmetric. All orbits in the region bounded by the manifolds from $F\rightarrow H\rightarrow E$ and $F\rightarrow G\rightarrow E$ represent bouncing cosmologies, the bounce occurring at the $h$-axis ($H=0$).}
    \label{skeleton}
\end{figure*}

The main focus of the analysis is to investigate whether different orbits inflate. The condition for a point in the phase space to be inflationary, using the deceleration parameter, $q$,  is given by
\begin{align}
	&q = -\frac{\dot{H}}{H^2}-1 < 0, \label{q.def}
\end{align}
which, using equation (11), in terms of the dimensionless variables, becomes
\begin{align}
	&q = \frac{3}{h^2}\left(h^2+\Phi^2-1\right)-1 < 0.\label{q.exp}
\end{align}
The results are presented in Table~\ref{ep_table}. We see a parameter space region allowing for two periods of inflation. One such example is Fig.~\ref{fig:1}. The expanding region of the phase diagram contains the saddle $H$ in the interior region and a sink $E$ on the boundary. Both points are inflationary (as are $G$ and $F$). At $H$, both fields contribute to inflation, while $E$ corresponds to standard single-field inflation. We can see that there exist orbits that approach a saddle and inflate for an arbitrarily long time, then approach the sink and inflate asymptotically again. That is shown in more detail in Fig.~\ref{skeleton}.

\begin{table*}[!ht]
	\centering
	\begin{tabular}{|c|c|c|c|c|c|}
		\hline 
		Label  & Existence & Sink & Saddle & Source & Inflation  \\\hline
		A & $\forall \lambda, k$  & $\kappa>0, \lambda<-\sqrt{6}$ & $\begin{array}{c}
			\kappa>0, \lambda>-\sqrt{6},\\
			\text{or} \\ \kappa<0, \lambda<-\sqrt{6}
		\end{array}$ & $\kappa<0, \lambda>-\sqrt{6}$ & N/A \\\hline 
		B & $\forall \lambda, k$  & $\kappa<0, \lambda>\sqrt{6}$ & $\begin{array}{c}
			\kappa<0, \lambda<\sqrt{6}, \\
			\text{or} \\  \kappa>0, \lambda>\sqrt{6}
		\end{array}$ & $\kappa>0, \lambda<\sqrt{6}$ & N/A \\\hline 
		C & $\forall \lambda, k$  & $\kappa>0, \lambda<\sqrt{6}$ & Same as $B$ & $\kappa<0, \lambda>\sqrt{6}$ & N/A \\ \hline 
		D & $\forall \lambda, k$  & $\kappa<0, \lambda>-\sqrt{6}$ & Same as $A$ & $\kappa>0, \lambda<-\sqrt{6}$ & N/A \\ \hline 
		$O_+$ & $\forall \lambda, k$  & $\kappa<0, \lambda<-\kappa$ & $\begin{array}{c}
			\kappa>0, \lambda<-\kappa\\ 
			\text{or} \\ \kappa<0, \lambda>-\kappa
		\end{array}$ & $\kappa>0, \lambda>-\kappa$ &  N/A \\\hline 
		$O_-$ & $\forall \lambda, k$  & $\kappa>0, \lambda>-\kappa$ & Same as $O_+$ & $\kappa<0, \lambda<-\kappa$ &  N/A \\\hline 
		$E$ & $\abs{\lambda}<\sqrt{6}$ & $\begin{array}{c}
			\abs{\lambda}<\sqrt{6},  \\ 
			\kappa \lambda +\lambda^2 -6 <0 
		\end{array}$ & $\begin{array}{c}
			\abs{\lambda}<\sqrt{6},  \\
			\kappa \lambda +\lambda^2 -6 >0 
		\end{array}$ & DNE & $\abs{\lambda}<\sqrt{2}$\\\hline
		$F$ & Same as $E$ & DNE & Same as $E$ &  $\begin{array}{c}
			\abs{\lambda}<\sqrt{6},  \\
			\kappa \lambda +\lambda^2 -6 <0 
		\end{array}$ & Same as $E$ \\\hline 
		$G$, $H$ & $\begin{array}{c}
			0<\kappa(\kappa+ \lambda), \\
			\kappa \lambda +\lambda^2 -6 <0
		\end{array}$ & DNE & $\begin{array}{c}
			\kappa>0,  
			\lambda( 6-\lambda^2) >\kappa (\kappa \lambda +2 \lambda^2-6) \\
			\text{or}\\
			\kappa<0,  
			\lambda( 6-\lambda^2) <\kappa (\kappa \lambda +2 \lambda^2-6)  
		\end{array}$ & DNE & $\frac{2}{3}<\frac{\kappa}{\kappa + \lambda}$ \\\hline 
	\end{tabular}
	
	\caption{Existence and stability of the equilibrium points of the reduced  dynamical system for the vector $(\Phi, h)$ given by \eqref{systA}-\eqref{systB}.}
	\label{ep_table}
\end{table*}

\subsection{Local effects of curvature}
\label{sec3}

Here we wish to simply determine the local stability of the inflationary equilibrium points $E$ and $H$. We note that, a global analysis can be achieved when $K \neq 0$ using appropriate bounded variables \cite{quintom6, PaperII}.  

Let us consider the case $K\neq0$ for the line element (\ref{sp.02}) and the Action Integral (\ref{sp.01}), which leads to the gravitational field equations
\begin{align}
	-&3H^2 +\frac{1}{2}\dot{\phi}^2 -\frac{1}{2}e^{\kappa\mkern1mu\phi}\dot{\psi}^2 +V\!\left( \phi \right) - {3K}{a^{-2}}=0 , \label{sp.04} \\
	2\dot{H} +&3H^2 +\frac{1}{2}\dot{\phi}^2 -\frac{1}{2}e^{\kappa\mkern1mu\phi}\dot{\psi}^2-V\!\left( \phi \right) + {K}{a^{-2}}=0 , \label{sp.05} \\
	\ddot{\phi}\, +&3H\dot{\phi} +\frac{1}{2}e^{\kappa\mkern1mu\phi}\dot{\psi}^2 +\partial_\phi V\!\left( \phi \right)=0 , \label{sp.06} \\
	\ddot{\psi}\, +&3H\dot{\psi} +\kappa\mkern1mu\dot{\phi}\mkern1mu\dot{\psi}=0. \label{sp.07}
\end{align}
Using the dimensionless variables \eqref{vars} and
\begin{equation}
	\Omega_K= {3}{a^{-2} \chi^{-2}}, \label{om}
\end{equation}
which satisfy
\begin{align}
	& h^2 + \eta^2 + K\Omega_{K}= \Psi + \Phi^2 = 1 , \label{rest1} 
\end{align}
and we utilize
\begin{equation}
	\frac{\dot{\chi}}{\chi^2}= -\frac{\Phi}{\sqrt{2}}\left( \kappa\eta^2 +\sqrt{6}h\Phi \right),
\end{equation}
from equations \eqref{sp.05}, \eqref{sp.06}, \eqref{sp.07} we obtain a system of equations for
$\{\Phi^\prime, h^\prime, \eta^\prime, \Psi^\prime,\Omega_K^\prime\}$ in terms of $\tau$ time.

For local stability, we can eliminate any two variables. To compare with the flat model, we will eliminate $\Psi$ and $\eta$ by substituting $\eta^2 = 1 - h^2 - K\Omega_{K}, \Psi = 1 - \Phi^2$. We then obtain the system:
\begin{align}
	&\Phi^\prime= -\frac{\left( 1-\Phi^2 \right)}{\sqrt{2}}\left( \sqrt{6}h\Phi + \kappa\left( 1 -h^2-K\Omega_K \right)+\lambda\right), \label{eq.(31)}\\ 
	&h^\prime= \sqrt{3}\left(\left(1-h^2-K\Omega_K\right)\left(1+\frac{\kappa h\Phi}{\sqrt{6}}\right)-\Phi^2\left(1-h^2\right)+\frac{1}{3}K\Omega_K\right), \label{eq.(32)}\\
	&\Omega_K^\prime= -2\Omega_K\left(\frac{h}{\sqrt{3}}-\frac{\Phi}{\sqrt{2}}\left(\sqrt{6}h\Phi+\kappa\left(1-h^2-K\Omega_K\right) \right) \right).
\end{align}

This subsection aims to see if the equilibrium points of interest, namely $E$ and $H$ (also $F$ and $G$), locally maintain their stability and inflationary nature. The corresponding points in this new system with coordinates $(\Phi,h,\Omega_K)$ are:
\begin{alignat*}{2}
	\tilde{E}&: \left(-\frac{\lambda}{\sqrt{6}}, 1, 0 \right), &&\qquad \tilde{G}: \left(\frac{\sqrt{6}}{\sqrt{\kappa^2 + \kappa \lambda +6}}, -\frac{\kappa + \lambda}{\sqrt{\kappa^2 + \kappa \lambda +6}}, 0 \right),  \\
	\tilde{F}&: \left(\frac{\lambda}{\sqrt{6}}, -1, 0 \right), &&\qquad \tilde{H}: \left(-\frac{\sqrt{6}}{\sqrt{\kappa^2 + \kappa \lambda +6}}, \frac{\kappa + \lambda}{\sqrt{\kappa^2 + \kappa \lambda +6}}, 0 \right).
\end{alignat*}
Note that the spatial curvature is zero at all of these equilibrium points. Similar to earlier, the condition for inflation becomes:
\begin{align}
    q = \frac{3}{h^2}\left(h^2+\Phi^2+\frac{2}{3}K\Omega_K-1\right)-1<0.
\end{align}
\noindent The additional eigenvalue (to the two presented earlier) for these equilibria are
\begin{align*}
\tilde{E}&: \left(\frac{\lambda^2}{\sqrt{3}}-\frac{2}{\sqrt{3}}\right),  &&\qquad 
\tilde{F}: &\left(-\frac{\lambda^2}{\sqrt{3}}+\frac{2}{\sqrt{3}}\right), \\
\tilde{G}&: \left(\frac{2\left(\kappa-2\lambda\right)}{\sqrt{3\left(\kappa^2+\kappa\lambda+6\right)}}\right), &&\qquad \tilde{H}: &\left(-\frac{2\left(\kappa-2\lambda\right)}{\sqrt{3\left(\kappa^2+\kappa\lambda+6\right)}}\right).
\end{align*}
\noindent We can now discuss the local stability of $E$ and $H$, which is essentially unchanged, so that $\tilde{E}$ is a sink and $\tilde{F}$ is a source for
\begin{align*}
    \lambda^2<2 && \text{and} && \kappa\lambda+\lambda^2-6<0
\end{align*}
\noindent whence both are inflationary. $\tilde{G}$ and $\tilde{H}$ are still saddle points, having two eigenvalues unchanged from the flat case, these being of opposite signs. These conditions are subsets of the flat space parameter range, so curvature does not prevent the possibility of a solution having two inflationary epochs.

\subsection{Bouncing solutions}
\label{bounce}

\begin{figure}[!ht]
    \centering
    \includegraphics[scale=0.55]{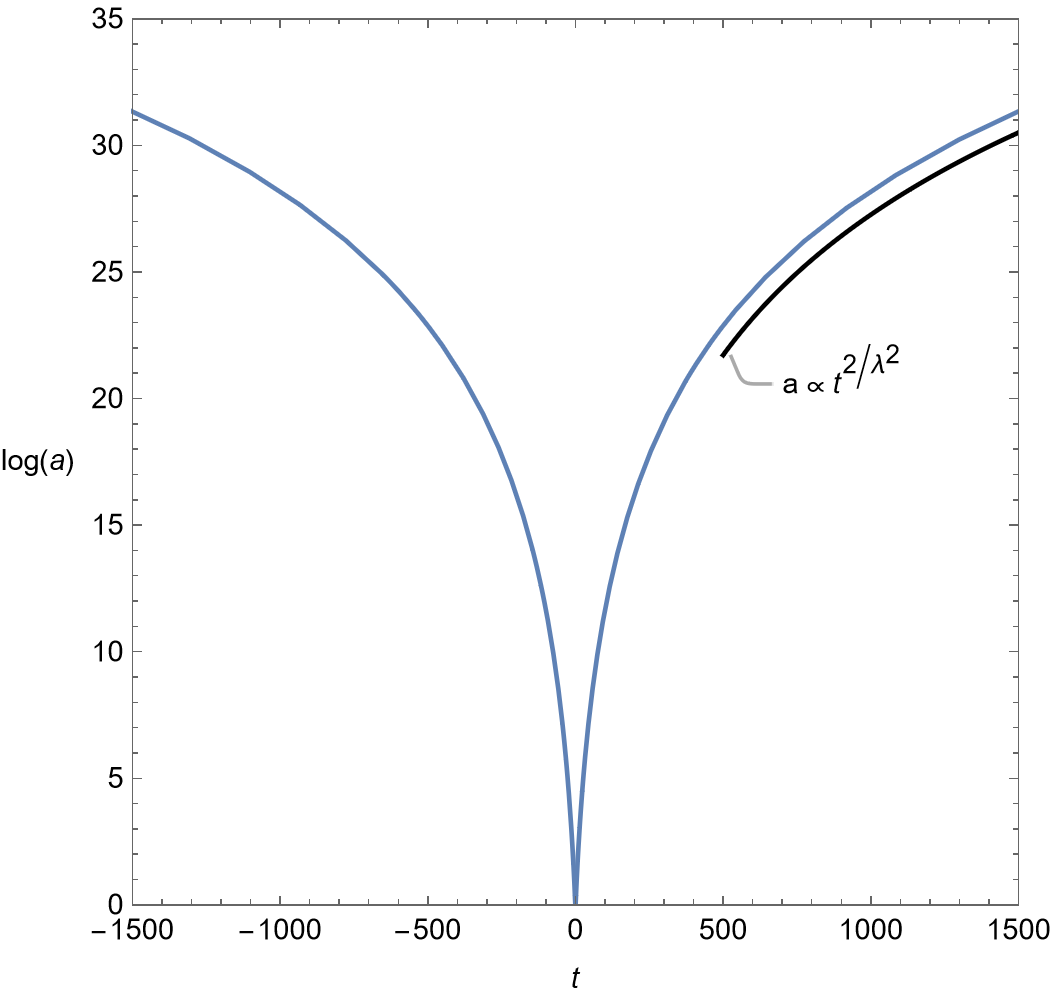}
    \caption{This figure shows the scale factor (logarithmically) vs proper time for the solution of \eqref{systA}-\eqref{systB} with initial conditions $\Phi=h=0$ at time $t=0$, for parameter values $\lambda=0.5,\kappa=1.5$. That corresponds to the green orbit shown in Fig.~\ref{skeleton} and represents a bouncing cosmology. The late-time accelerating expansion is $a\propto t^{2/\lambda^2}$.}
    \label{bouncing}
\end{figure}

In addition to numerically integrating the main differential equations \eqref{systA}-\eqref{systB}, we can also solve for the relevant cosmological quantities by including ancillary equations that decouple from the system. For example, the Hubble rate $H(\tau)$ can be determined from \eqref{hubble}, while \eqref{chi_de} allows us to solve for $\chi(\tau)$:
\begin{align}
    & H^\prime=\chi(\tau)\left( 1-h^2-\Phi^2 \right); 
    \chi^\prime=-\chi\frac{\Phi}{\sqrt{2}} \left( \kappa\left( 1-h^2 \right) +\sqrt{6}h\Phi \right). \label{chi_de2}
\end{align}
\noindent The scale factor for the cosmological evolution can be found simply by integrating $h(\tau)$, since
\begin{equation}
    h=\sqrt{3}\,\frac{d}{d\tau}\ln(a),\label{eq-a}
\end{equation}
\noindent and we can compare all functions of $\tau$ with the proper time $t$ using $d\tau=\chi\,dt$.

Fig.~\ref{skeleton} shows a skeleton of the phase portrait for the dynamical system \eqref{systA}-\eqref{systB} for the parameter values $\lambda=0.5,\kappa=1.5$. Shown in green is the solution which passes through the origin $\Phi=h=0$ at $t=0$. An analytical approximation for this solution can be obtained. This solution is completely time-symmetric about the origin, and represents a bouncing cosmology, with $H=0$ at $t=0$, $H>0$ and for $t>0$, and $H<0$ for $t<0$.  All orbits in the region bounded by the stable manifolds connecting $H$ and $G$ to $F$ and the unstable manifolds connecting $H$ and $G$ to $E$ correspond to bouncing cosmologies. The bounce occurs where these orbits cross the $h$-axis. Fig.~\ref{bouncing} shows the scale factor $\ln(a)$ vs proper time for the time-symmetric solution (the green orbit in Fig.~\ref{skeleton}). Note that at $t=0$ where $\dot{a}=0$ , $\ln(a)=0$ (i.e., $a=1$), so the bounce is regular ($a$ never approaches 0).
\noindent
The asymptotic behaviour of the expansion can also be derived for solutions for which $(\Phi,h)\rightarrow E$ as $t\rightarrow\infty$. Recalling the deceleration $q$ defined by \eqref{q.def}, it can be seen that as $q$ approaches a constant at late times; the Hubble rate is $H\sim t^{-1}/(1+q)$ to leading order, which has expansion $a\sim t^{\left(1+q\right)^{-1}}$. From \eqref{q.exp}, we see that the deceleration at the equilibrium point $E$ is $1+q(E)=\lambda^2/2$. Thus for solutions of \eqref{systA}-\eqref{systB} that approach $E$ at late times, the asymptotic behaviour of the expansion is $a\sim t^{2/\lambda^2}$.

That is one of many possible behaviours of orbits asymptotic to $F$ at early times. For example, in a different region of parameter space, the stable manifold of $H$ connects to both $B$ and ${O_{+}}$, while the unstable manifold of $G$ connects to $C$ and ${O_{-}}$. In this case, all orbits from $F$ are in the basin of attraction of ${O_{-}}$, while all orbits in the basin of attraction of $E$ have ${O_{+}}$ as their early-time past, and the solution that passes through the origin goes from ${O_{+}}$ at early times to ${O_{-}}$ at late times (e.g., see Fig.~\ref{fig:1a}). More details will be provided in \cite{PaperII}. In the bouncing case, when the origin is in the basis of attraction of $E$, then $h\rightarrow 1^-$ as $t\rightarrow\infty$. In the latter case, we have $h\rightarrow0^+$ as $t\rightarrow\infty$.

\subsection{Analysis of linear cosmological perturbations}
\label{linear-perts}

The action \eqref{sp.01} is a subcase of the N-field chiral action \cite{Chervon:2013btx}
   \begin{equation}
	S=\int\sqrt{-g}dx^{4}\left(  R-\frac{1}{2}h_{A B}g^{\mu\nu}\varphi^A_{, \mu}\varphi^B_{,\nu} -W\left( \varphi^C\right)  \right), \label{chiral}
\end{equation}
where $\varphi$ is a multiplet of chiral fields lying in a chiral space with metric $h_{A B}$,  by specifying  $\varphi:=(\varphi^1, \varphi^2)=(\phi, \psi)$ and 
$h_{11}=1, h_{1 2}=h_{21}=0, h_{2 2}=-e^{\kappa\mkern1mu\phi}$ and $W\left( \varphi^C\right) =V(\phi)=V_0 e^{\lambda \phi}$. 

For the analysis of linear cosmological perturbations we consider perturbed spacetime around the spatially flat FLRW space written as follows \cite{amebook}: 
\begin{align}
ds^{2}& =a^{2}\left(  \sigma\right) 
\Big(  -\left(  1+2\mathcal{S}\left(
\sigma,x,y,z\right)  \right) 
d\sigma^{2} \nonumber \\
& +\left(  1-2\mathcal{S}\left(
\sigma,x,y,z\right)  \right) \gamma_{i j}
dx^{i}dx^{j}\Big) 
,\label{pr.01}%
\end{align}
where $\sigma$ is the conformal time given by 
 \begin{align}
     & \frac{d}{d\sigma} =a \frac{d}{dt}, \; \frac{d^2}{d\sigma^2}=a^2H\frac{d}{dt}+a^2\frac{d^2}{dt^2}, \label{1d}
 \end{align} 
The function $\mathcal{S}\left(  \sigma,x,y,z\right)  $ includes the perturbation terms. The scalar fields $\varphi^A$  are perturbed
as ${\tilde{\varphi}}^A= {\varphi}^A (t)+\delta\varphi^A (x,t)$ and the metric in the chiral space is perturbed as \newline 
$\tilde{h}_{A B}(\tilde{\varphi}^C)={h}_{A B}(\tilde{\varphi}^C)= h_{A B}(\varphi^C(t))+ \partial_C h_{A B}\delta\varphi^{C}(x,t)$, where $h_{A B}$ and $\varphi^C$, $A, B, C =1 \ldots N$, are the background quantities.  

Moreover, we consider the linear perturbation theory in the Newtonian gauge. In these gauge, the perturbation of the background equations for the FLRW metric are given by  \cite{Chervon:2013btx}
\begin{align}
& 3 \left({\mathcal{H}}^2+K\right)=V_0 a^2 e^{\lambda  \phi}-\frac{1}{2} e^{\kappa  \phi} {\psi^{\prime}}^2+\frac{1}{2} {\phi^{\prime}}^2, \\
   & 2 \mathcal{H}^{\prime}+ \mathcal{H}^2+K= V_0 a^2 e^{\lambda  \phi} +\frac{1}{2} e^{\kappa  \phi} {\psi  ^{\prime}}^2-\frac{1}{2} {\phi ^{\prime}}^2,\\
   & \phi^{\prime\prime}=- \lambda V_0 e^{\lambda  \phi} a^2  -2
   \mathcal{H} \phi^\prime-\frac{1}{2} \kappa  e^{\kappa  \phi } {\psi ^\prime}^2, \label{eq.(42)}\\
   & \psi^{\prime\prime}= -2 \mathcal{H} \psi ^\prime -\kappa  \psi^\prime \phi^\prime. 
\end{align}

\noindent 
where the prime and ${}_{,\sigma}$ indicates a derivative with respect to the conformal time $\sigma$, and 
$\mathcal{H}=\frac{{a}^{\prime}}{a}= a H$ is the Hubble parameter in the new frame, $i,j,k=1,2,3$ and   $K=0,1,-1$.

The perturbed Einstein equations are specialised to 
\begin{align}
    & \nabla^2 \mathcal{S} - 3 \mathcal{H} \mathcal{S}^{\prime} - \mathcal{S} \left(2 \mathcal{H}^2 + \mathcal{H}^{\prime}\right) + 4 K \mathcal{S} \nonumber \\
    &=\frac{1}{2}\left\{{\varphi^{\prime}} \phi^{\prime} -e^{\kappa\phi} {\xi}^{\prime} \psi^{\prime} -\frac{1}{2}\kappa e^{\kappa\phi}  {\psi^{\prime}}^2 {\varphi} + a^2 \lambda V_0 e^{\lambda\phi} {\varphi}\right\},\label{perts1}\\
    & \left(\mathcal{H} \mathcal{S} + \mathcal{S}^{\prime}\right)_{,i}= \frac{1}{2} \varphi_{,i} \phi^{\prime} - \frac{1}{2} e^{\kappa \phi} {\xi}_{,i}{\psi}^{\prime}, \label{perts2}\\
    & \mathcal{S}^{\prime \prime} + 3 \mathcal{H} \mathcal{S}^{\prime} + \mathcal{S} \left(2 \mathcal{H}^2 + \mathcal{H}^{\prime}\right) \nonumber \\
    &+\frac{1}{2}\left\{{\varphi^{\prime}} \phi^{\prime} -e^{\kappa\phi} {\xi}^{\prime} \psi^{\prime} -\frac{1}{2}\kappa e^{\kappa\phi}  {\psi^{\prime}}^2 {\varphi} -a^2 \lambda V_0 e^{\lambda\phi} \varphi\right\}, \label{perts3}
\end{align}
where we designated $\varphi=\delta\phi$ and $\xi=\delta\psi$. 

The master equation for $\mathcal{S}$ is obtained by adding \eqref{perts1} and \eqref{perts3} which leads to 
\begin{align}
     & \nabla^2 \mathcal{S}  + \mathcal{S}^{\prime \prime} + 4 K \mathcal{S} ={\varphi^{\prime}} \phi^{\prime} -e^{\kappa\phi} {\xi}^{\prime} \psi^{\prime} -\frac{1}{2}\kappa e^{\kappa\phi}  {\psi^{\prime}}^2 {\varphi}. \label{eq.(48)}
\end{align}
In equation \eqref{perts2}, the  expressions $\phi^{\prime}$ and ${\psi}^{\prime}$ are quantities independent of  spatial coordinates, so by integration, we have that 
\begin{align}
\mathcal{H} \mathcal{S} + \mathcal{S}^{\prime}= \frac{1}{2} \varphi \phi^{\prime} - \frac{1}{2} e^{\kappa \phi} {\xi}{\psi}^{\prime}. \label{eq.(49)}
\end{align}
Now let us decompose the perturbation of the gravitation field $\mathcal{S}$ into two parts:
\begin{equation}
    \mathcal{S}(x,t)= \mathcal{S}_{\phi} + \mathcal{S}_{\psi}=\Xi +\Sigma \label{eq.(50)}
\end{equation}
where $\Xi$ is responsible for the perturbations coming from the field $\phi$,  $\Sigma$ is responsible for the perturbations coming from the field $\psi$, and decompose the scalar product in the target space as 
\begin{equation}
    h_{A B} \varphi^A_\mu  \varphi^B_\nu= (\phi^{\prime})^2 - e^{\kappa \phi}  (\psi^{\prime})^2. \label{eq.(51)}
\end{equation}
The decomposition \eqref{eq.(50)}, \eqref{eq.(51)} allows separate the perturbations $\Xi$ and $\Sigma$ in the perturbed Einstein equations  (see Equations (50)-(53) of \cite{Chervon:2013btx}). 
Equations \eqref{eq.(48)} and \eqref{eq.(49)} are now expressed as 
\begin{align}
    & \Xi^{\prime \prime} +\nabla^2 \Xi + 4 K \Xi = \phi^{\prime} \varphi^{\prime},\\
    & \Sigma^{\prime \prime} +\nabla^2 \Sigma + 4 K \Sigma = -e^{\kappa\phi} {\xi}^{\prime} \psi^{\prime} -\frac{1}{2}\kappa e^{\kappa\phi}  {\psi^{\prime}}^2 {\varphi}, \\
    & \mathcal{H} \Xi + \Xi^{\prime}= \frac{1}{2} \varphi \phi^{\prime},\\
    & \mathcal{H} \Sigma + \Sigma^{\prime}= - \frac{1}{2} e^{\kappa \phi} {\xi}{\psi}^{\prime}.
\end{align}
The total equation of the state of the matter fields (for $K=0$) satisfy 
\begin{equation}
    \frac{ {\mathcal{H}}^{\prime}}{\mathcal{H}^2}= -\frac{1}{2}\left(1+3w_{\text{eff}}\right)= -q,
\end{equation}
where $q$ is the deceleration parameter given by \eqref{q.exp}.

Assuming $K=0$, to obtain a dynamical system that describes the evolution of the perturbations, we first introduce Cartesian spatial coordinates and make the Fourier transform of the perturbation variables
\begin{equation}
\left( \varphi, \xi,   \Xi\right) =\int \left( \varphi_k, \xi_k, \Xi_k\right) e^{- i k \cdot \mathbf{x}}dxdydz, 
\end{equation} 
whence,  equations \eqref{eq.(48)} and \eqref{eq.(49)} become 
\begin{align}
    & \Xi^{\prime \prime} -k^2 \Xi  = \phi^{\prime} \varphi^{\prime}, \label{eq.(63)}\\
    & \Sigma^{\prime \prime} -k^2 \Sigma  = -e^{\kappa\phi} {\xi}^{\prime} \psi^{\prime} -\frac{1}{2}\kappa e^{\kappa\phi}  {\psi^{\prime}}^2 {\varphi}, \label{eq.(64)}
\\
    & \mathcal{H} \Xi + \Xi^{\prime}= \frac{1}{2} \varphi \phi^{\prime}, \label{eq.(65)}\\
    & \mathcal{H} \Sigma + \Sigma^{\prime}= - \frac{1}{2} e^{\kappa \phi} {\xi}{\psi}^{\prime} \label{eq.(66)}.
\end{align}
The last equations give definitions of the perturbations $\varphi=\delta\phi$ and $\xi=\delta\psi$ and we introduce a change of variables $ \xi=\zeta e^{-\frac{\kappa}{2}\phi}$.

Passing to the time derivative 
 $d\tau=\chi d t$ we have 
  \begin{align}
     & \frac{d}{d t} =\chi \frac{d}{d\tau}, \nonumber \\
     & \frac{d^2}{dt^2}=\left[-\frac{\Phi}{\sqrt{2}} \left( \kappa\left( 1-h^2 \right) +\sqrt{6}h\Phi \right)\right]\chi^2\frac{d}{d\tau}+\chi^2\frac{d^2}{d\tau^2}. \label{derivatives}
 \end{align}
In the new variables, eqs. \eqref{eq.(65)} and \eqref{eq.(66)}
become
\begin{align}
       & \frac{h \Xi }{\sqrt{3}}+\Xi^{\prime}-\frac{\varphi \Phi }{\sqrt{2}}=0, \label{eq.(74)}\\
   & \frac{1}{6} \left(3 \zeta  \sqrt{2-2 h^2}+2 \sqrt{3} h \Sigma \right)+\Sigma^{\prime}=0,  \label{eq.(75)}
\end{align}
where the prime means derivative with respect to $\tau$. 
Then, we define $Z=k^2(a\chi)^{-2}$,
such that 
\begin{align}
    Z^{\prime}=Z \left(\sqrt{2} \Phi  \left(-\kappa  h^2+\sqrt{6} h \Phi +\kappa \right)-\frac{2 h}{\sqrt{3}}\right). \label{Eq.(76)}
\end{align}
\noindent
Equations  \eqref{eq.(74)} and \eqref{eq.(75)} gives the definitions 
\begin{align}
   &\varphi= \frac{\sqrt{2} \left(\sqrt{3} h\Xi +3 \Xi ^{\prime}\right)}{3 \Phi }, \quad \zeta=-\frac{\sqrt{2}
   \left(\sqrt{3} h\Sigma +3 \Sigma ^{\prime}\right)}{3 \sqrt{1-h^2}}\label{pert-scalars}
\end{align}
for the perturbed scalar fields. 

In the new variables equations \eqref{eq.(63)} and \eqref{eq.(64)} are given by 
\begin{align}
   & - \Xi Z+\left(\frac{\left(h^2-1\right) \kappa  \Phi }{\sqrt{2}}+\frac{h
   \left(1-3 \Phi ^2\right)}{\sqrt{3}}\right) \Xi^{\prime} +\Xi^{\prime \prime}-\sqrt{2} \Phi  \varphi^{\prime}=0,\\
   & - \Sigma Z+\sqrt{2-2 h^2} \zeta^{\prime} +\left(\frac{\left(h^2-1\right) \kappa  \Phi
   }{\sqrt{2}}+\frac{h \left(1-3 \Phi ^2\right)}{\sqrt{3}}\right) \Sigma^{\prime}\nonumber\\
   & +\kappa  \left(-\zeta  \sqrt{1-h^2}
   \Phi +(1-h^2)\varphi\right)+\Sigma^{\prime \prime}=0. \label{evolve_Sigma2}
\end{align}
From the expression $\varphi$ in \eqref{pert-scalars} and the equations \eqref{systA}-\eqref{systB}  we obtain
\begin{align}
   & \Xi^{\prime\prime}= \frac{1}{3} \Xi  \left(\frac{\sqrt{6} \kappa  h \left(h^2-1\right)+\left(\Phi ^2-1\right) \left(\sqrt{6} \lambda  h+6 \Phi \right)}{\Phi}-3 Z\right) \nonumber\\
   & +\Xi ^{\prime} \left(-\frac{\kappa  \left(h^2-1\right)
   \left(\Phi ^2-2\right)}{\sqrt{2} \Phi }+\frac{h \left(3 \Phi^2-7\right)}{\sqrt{3}}+\frac{\sqrt{2} \lambda  \left(\Phi ^2-1\right)}{\Phi}\right).\label{eq.(70)}
\end{align}
\noindent
We write ${\Xi}=f_1+i f_2$, where $f_1$ and $f_2$ are the real and imaginary parts of the ${\Xi}$, respectively.  The resulting equations \eqref{eq.(70)}  have the same structure for $f_1$ and $f_2$. Therefore, we omit the sub-index. We denote $f=r\cos\theta$ and $f^{\prime}=r\sin\theta$ and
$f^{\prime}=f  \tan\theta$. Hence, equation \eqref{eq.(70)} is a two-degree equation that can be converted to one degree by using the trigonometric relations, say, 
\begin{align}
 &   \theta^{\prime} =-\sin^2\theta-P \sin\theta \cos\theta -Q\cos^2\theta,\label{Eq.(73)}
\end{align}
where
\begin{align}
&P=  \left(\frac{\kappa  \left(1-h^2\right)
   \left(\Phi ^2-2\right)}{\sqrt{2} \Phi }+\frac{h \left(3 \Phi^2-7\right)}{\sqrt{3}}+\frac{\sqrt{2} \lambda  \left(\Phi ^2-1\right)}{\Phi}\right),\\
&Q=\frac{1}{3}\left(\frac{\sqrt{6} \kappa  h \left(h^2-1\right)+\left(\Phi ^2-1\right) \left(\sqrt{6} \lambda  h+6 \Phi \right)}{\Phi}-3 Z\right).
\end{align} 
Taking derivatives of \eqref{pert-scalars} with respect to  $\tau$ using equations \eqref{systA}, \eqref{systB} and \eqref{eq.(75)}, and defining $\omega=\zeta  \sqrt{1- h^2}= e^{\kappa \phi} \delta\psi \frac{d{\psi}}{d\tau}$ we deduce the auxiliary equations 
\begin{align} 
  & \varphi^{\prime}= \frac{\sqrt{2} \Xi   \left(h ^2+\Phi  ^2-Z -1\right)}{\Phi  }-\frac{\varphi  \left(\kappa(1- h ^2)+\sqrt{6} h  \Phi -\lambda  \Phi  ^2+\lambda \right)}{\sqrt{2} \Phi  }, \label{evolve_v}\\
  & \omega ' =-\frac{\kappa  \left(h ^2-1\right) \Phi   \omega  }{\sqrt{2}}+\sqrt{3} h \left(\Phi  ^2-2\right) \omega  +\sqrt{2} \kappa  \left(h ^2-1\right) \varphi \nonumber\\
  & -\sqrt{2} \Sigma   \left(h ^2+\Phi  ^2-Z -1\right),\label{evolve_omega}
\\
  & \Sigma^{\prime}=  -\frac{h  \Sigma  }{\sqrt{3}}-\frac{\omega  }{\sqrt{2}}, \label{evolve_Sigma}\\
  & \Xi^{\prime}= \frac{\varphi \Phi }{\sqrt{2}} -\frac{h \Xi }{\sqrt{3}}. \label{evolve_Xi}
\end{align}

\subsubsection{Extended phase space}

The state space $S$ of the system \eqref{systA} and \eqref{systB} augmented with the perturbation equations \eqref{Eq.(76)} and \eqref{Eq.(73)} admits the product structure $S = B \times P$. $B$ is the background state space, which describes the dynamics of an FLRW background, and $P$ is the perturbation state space, containing gauge invariant variables that describe linear cosmological perturbations,  with $\{\Phi, h\}\subset B$ and $\{Z,\theta\}\subset P$ as in \cite{amebook, Wainwright:2005,am0, Basilakos:2019dof, Alho:2019jho, Alho:2020cdg}. 

Defining the compact variable 
\begin{align}
    \bar{Z}=(1+Z)^{-1} Z = {k^2}/(k^2+(a\chi)^2),\quad 
    Z=(1-\bar{Z})^{-1} {\bar{Z}}, 
\end{align} and changing time from   $\tau$ to $\bar{\tau}$ through the formula 
\begin{align}
{d\bar{\tau}}/{d\tau}=(1-\bar{Z})^{-1}=1+Z,
\end{align}
we obtain 
\begin{align}
& \Phi^{\prime} =   \frac{1}{2} \left(\Phi ^2-1\right) \left(\bar{Z}-1\right) \left(\sqrt{2}
   \kappa  h^2-2 \sqrt{3} h \Phi -\sqrt{2} (\kappa +\lambda )\right), \label{final-system1}\\
& h^{\prime}=\frac{1}{2} \left(h^2-1\right) \left(\bar{Z}-1\right) \left(\sqrt{2}\kappa  h \Phi -2 \sqrt{3} \left(\Phi ^2-1\right)\right),\label{final-system2}
\\
&{\bar{Z}}^{\prime}=\frac{1}{3} \left(\bar{Z} -1\right)^2 \bar{Z}  \left(2 \sqrt{3} h  \left(3 \Phi
    ^2-1\right)-3 \sqrt{2} \kappa  \left(h ^2-1\right) \Phi  \right),
  \label{final-system3}\\
&{\theta}^{\prime}= \Phi  \left(\bar{Z}-1\right) \left(\frac{\sin (2
   \theta ) \left(-\kappa  h^2+\kappa +2 \lambda \right)}{2
   \sqrt{2}}+\sqrt{\frac{2}{3}} \lambda  h \cos ^2(\theta )\right)\nonumber\\
   & +\frac{\left(\bar{Z}-1\right) \left(\frac{\sin (2 \theta ) \left(\kappa
    h^2-\kappa -\lambda \right)}{\sqrt{2}}+\sqrt{\frac{2}{3}} h \cos ^2(\theta
   ) \left(\kappa  h^2-\kappa -\lambda \right)\right)}{\Phi }\nonumber\\
   & +\Phi ^2  \left(\bar{Z}-1\right) \left(\frac{\sqrt{3}}{2} h \sin (2\theta)+2 \cos^2 (\theta )\right)\nonumber\\
   &-\frac{7 h \left(\bar{Z}-1\right) \sin (2\theta )}{2\sqrt{3}}  +\frac{1}{2} \left(\left(3-2 \bar{Z}\right)
   \cos (2 \theta )+1\right).\label{final-system4}
\end{align}
defined in $(\Phi, h,\bar{Z},\theta) \in [-1,1]\times [-1,1] \times [0,1] \times[0,2\pi]$. Equation \eqref{final-system4} is periodic in $\theta$ of period $\pi$ . In table \ref{tab:perts} we present the equilibrium points $(\Phi,h, \bar{Z}, \theta)$ of system \eqref{final-system1}, \eqref{final-system2}, \eqref{final-system3}
and \eqref{final-system4}, modulo $n\pi, \; n\in \mathbb{Z}$:
\begin{align*}
A&: (1, 1, 0, 0),  \qquad A^*: \left(1, 1, 0, \tan ^{-1}\left(\frac{4}{\sqrt{3}}\right)\right), \\
B&: (-1, 1, 0, 0),  \qquad B^*: \left(-1, 1, 0, \tan ^{-1}\left(\frac{4}{\sqrt{3}}\right)\right),  \\
C&: (1, -1, 0, 0),  \qquad C^*: \left(1,-1, 0, -\tan ^{-1}\left(\frac{4}{\sqrt{3}}\right)\right), \\
D&: (-1, -1, 0, 0),   \qquad D^*: \left(-1,-1, 0, -\tan ^{-1}\left(\frac{4}{\sqrt{3}}\right)\right), \\
O_{\pm}&: (\pm1, 0, 0, 0),   \qquad O^*_{\pm}: \left(\pm1,0, 0, \pm\tan ^{-1}\left(\frac{\kappa }{\sqrt{2}}\right)\right), \\
E &: \left(-\frac{\lambda}{\sqrt{6}}, 1, 0, 0 \right), 	  \qquad E^*: \left(-\frac{\lambda}{\sqrt{6}},  1, 0,  \tan ^{-1}\left(\frac{\lambda ^2+2}{2 \sqrt{3}}\right)\right),  \\
F &: \left(\frac{\lambda}{\sqrt{6}},  -1, 0, 0 \right),	 \qquad F^*: \left( \frac{\lambda}{\sqrt{6}}, - 1, 0,  -\tan ^{-1}\left(\frac{\lambda ^2+2}{2 \sqrt{3}}\right)\right), 
\end{align*}

\begin{table*}[!ht]
	\centering
\small\setlength{\tabcolsep}{5pt}
	\begin{tabular}{|c|c|c|c|c|c|c|c|c|c|}
		\hline 
Label   & $\theta$ & $k_1$ & $k_2$ & $k_3$ & $k_4$ & Stability  \\ \hline
$ A$  &  $0$ &$ \frac{4}{\sqrt{3}}$ &$ \frac{4}{\sqrt{3}} $ & $-\sqrt{2} \kappa $ &$ \sqrt{2} \lambda+2 \sqrt{3} $& source for $\kappa<0, \lambda>-\sqrt{6}$, or saddle \\
$ A^*$  & $\tan ^{-1}\left(\frac{4}{\sqrt{3}}\right)$ & $-\frac{4}{\sqrt{3}}$ &   $ \frac{4}{\sqrt{3}}$  &  $-\sqrt{2} \kappa$ & $ \sqrt{2} \lambda +2 \sqrt{3}$ & saddle \\
$ \bar{A}$ & $-\frac{\pi}{2}$ & $-\frac{4}{\sqrt{3}}$ & $0$ & $-\sqrt{2} \kappa$ & $\sqrt{2} \lambda +2 \sqrt{3}$ & sink for $\kappa >0, \lambda <-\sqrt{6}$, or saddle \\
$B$  &  $0$ & $\frac{4}{\sqrt{3}} $& $\frac{4}{\sqrt{3}}$ & $\sqrt{2} \kappa$  & $2 \sqrt{3}-\sqrt{2} \lambda$  &  source for $\kappa>0, \lambda<\sqrt{6}$, or saddle \\
$B^*$  & $\tan ^{-1}\left(\frac{4}{\sqrt{3}}\right)$ & $-\frac{4}{\sqrt{3}}$ & $ \frac{4}{\sqrt{3}}$ & $\sqrt{2} \kappa $ & $2 \sqrt{3}-\sqrt{2} \lambda $ & saddle \\
$\bar{B}$ & $-\frac{\pi}{2}$ & $-\frac{4}{\sqrt{3}}$ & $0$ & $\sqrt{2} \kappa $ & $2 \sqrt{3}-\sqrt{2} \lambda$& sink for $\kappa <0, \lambda >\sqrt{6}$, or saddle \\
$C$ &  $0$ & $-\frac{4}{\sqrt{3}}$ & $-\frac{4}{\sqrt{3}}$ &$ -\sqrt{2} \kappa $ & $\sqrt{2}  \lambda -2 \sqrt{3} $& sink for  $\kappa>0, \lambda<\sqrt{6}$, or saddle \\
$C^*$  & $ -\tan ^{-1}\left(\frac{4}{\sqrt{3}}\right)$ & $-\frac{4}{\sqrt{3}}$ & $ \frac{4}{\sqrt{3}} $& $-\sqrt{2} \kappa $ & $ \sqrt{2} \lambda -2 \sqrt{3} $ & saddle \\
$ \bar{C}$ & $-\frac{\pi}{2}$ & $\frac{4}{\sqrt{3}}$ & $0$ & $-\sqrt{2} \kappa$ & $\sqrt{2} \lambda -2 \sqrt{3}$ & source for $\kappa <0, \lambda >\sqrt{6}$, or saddle \\
$D$   &  $0$ & $-\frac{4}{\sqrt{3}}$ & $-\frac{4}{\sqrt{3}}$ & $\sqrt{2} \kappa$  & $-\sqrt{2} \lambda -2 \sqrt{3} $& sink for $\kappa<0, \lambda>-\sqrt{6}$, or saddle \\
$D^*$  & $-\tan ^{-1}\left(\frac{4}{\sqrt{3}}\right)$ & $-\frac{4}{\sqrt{3}}$ &  $\frac{4}{\sqrt{3}}$ & $\sqrt{2} \kappa $ & $-\sqrt{2} \lambda -2 \sqrt{3}$ & saddle \\
$ \bar{D}$ & $-\frac{\pi}{2}$ & $\frac{4}{\sqrt{3}}$ & $0$ & $\sqrt{2} \kappa$ & $-\sqrt{2} \lambda -2 \sqrt{3}$ & source for $\kappa >0, \lambda <-\sqrt{6}$, or saddle \\
$O_{+}$  &  $0$ & $\frac{\kappa }{\sqrt{2}}$ & $\frac{\kappa }{\sqrt{2}}$ & $\sqrt{2} \kappa $ & $\sqrt{2} (\kappa +\lambda ) $& sink for $\kappa<0, \lambda<-\kappa$ \\
      & & & & & & source for $\kappa>0, \lambda>-\kappa$ \\   
       &&&&&& saddle otherwise \\
$O_{+}^*$  & $\tan ^{-1}\left(\frac{\kappa }{\sqrt{2}}\right)$ & $\frac{\kappa }{\sqrt{2}}$ & $-\frac{\kappa }{\sqrt{2}}$ & $\sqrt{2} \kappa$  &$ \sqrt{2} (\kappa +\lambda ) $ & saddle\\
$ \bar{O}_+$ & $-\frac{\pi}{2}$ & $\frac{\kappa }{\sqrt{2}}$ & $0$ & $-\sqrt{2} \kappa$ & $\sqrt{2} (\kappa +\lambda )$ & saddle \\
$O_{-}$   &$0$ &  $-\frac{\kappa }{\sqrt{2}}$ & $-\frac{\kappa }{\sqrt{2}}$  & $-\sqrt{2} \kappa$ & $-\sqrt{2} (\kappa +\lambda )$ & sink for $\kappa>0, \lambda>-\kappa$ \\
      & & & & & & source for $\kappa<0, \lambda<-\kappa$   \\  
       &&&&&& saddle otherwise \\
$O_{-}^*$  & $-\tan ^{ -1}\left(\frac{\kappa }{\sqrt{2}}\right)$ &  $ -\frac{\kappa }{\sqrt{2}} $& $\frac{\kappa }{\sqrt{2}}$ & $-\sqrt{2} \kappa $ & $ -\sqrt{2} (\kappa +\lambda )$ & saddle \\
$ \bar{O}_-$ & $-\frac{\pi}{2}$ & $-\frac{\kappa }{\sqrt{2}}$ & $0$ & $\sqrt{2} \kappa$ & $\sqrt{2} (\kappa +\lambda )$ & saddle \\
$E$  &  $0$& $\frac{\lambda ^2-2}{\sqrt{3}}$ & $\frac{\lambda ^2+2}{2 \sqrt{3}} $ & $\frac{\lambda ^2-6}{2 \sqrt{3}}$ & $\frac{\lambda  (\kappa +\lambda)-6}{\sqrt{3}}$ & saddle \\
$E^* $  & $  \tan ^{-1}\left(\frac{\lambda ^2+2}{2 \sqrt{3}}\right) $ & $\frac{\lambda ^2-2}{\sqrt{3}}$ &$ -\frac{\lambda ^2+2}{2\sqrt{3}}$&    $\frac{\lambda ^2-6}{2 \sqrt{3}} $ &$ \frac{\lambda  (\kappa +\lambda )-6}{\sqrt{3}} $&  sink for $ \abs{\lambda}<\sqrt{2},   \kappa \lambda +\lambda^2 -6 <0$\\
		  & & & & & & saddle otherwise\\
$\bar{E}$ & $-\frac{\pi}{2}$ & $-\frac{\lambda ^2-2}{\sqrt{3}}$ & $0$ & $\frac{\lambda ^2-6}{2 \sqrt{3}}$ & $\frac{\lambda  (\kappa +\lambda )-6}{\sqrt{3}}$ &  sink for $\sqrt{2}<\abs{\lambda}<\sqrt{6},   \kappa \lambda +\lambda^2 -6 <0$\\
		  & & & & & & saddle otherwise\\
$F$  &  $0$ & $-\frac{\lambda ^2-2}{\sqrt{3}}$ & $-\frac{\lambda ^2+2}{2 \sqrt{3}}$  &$ -\frac{\lambda ^2-6}{2 \sqrt{3}}$  & $\frac{6-\lambda (\kappa +\lambda )}{\sqrt{3}}$ & saddle \\
$F^*$ &$-\tan ^{-1}\left(\frac{\lambda ^2+2}{2 \sqrt{3}}\right)$ & $-\frac{\lambda ^2-2}{\sqrt{3}}$ & $\frac{\lambda ^2+2}{2 \sqrt{3}}$  &$ -\frac{\lambda ^2-6}{2 \sqrt{3}}$  & $\frac{6-\lambda (\kappa +\lambda )}{\sqrt{3}}$ & source for $  \abs{\lambda}<\sqrt{2},  	\kappa \lambda +\lambda^2 -6 <0$\\
	 & & & & & & saddle otherwise\\
$\bar{F}$ & $-\frac{\pi}{2}$ & $\frac{\lambda ^2-2}{\sqrt{3}}$ & $0$ & $-\frac{\lambda ^2-6}{2 \sqrt{3}}$ & $\frac{6-\lambda  (\kappa +\lambda )}{\sqrt{3}}$ & source for $\sqrt{2}<\abs{\lambda}<\sqrt{6},   \kappa \lambda +\lambda^2 -6 <0$\\
		  & & & & & & saddle otherwise\\	 
$G$  &  $0$ & $\frac{4 \lambda -2 \kappa }{\sqrt{3}
   \sqrt{\kappa  (\kappa +\lambda )+6}}$ & $\frac{\kappa +4 \lambda }{\sqrt{3} \sqrt{\kappa 
   (\kappa +\lambda )+6}}$ &$ \delta_+ $& $\delta_-$ & saddle \\
$G^*$  &  $  \tan ^{-1}\left(\frac{\kappa +4 \lambda }{\sqrt{3} \sqrt{\kappa  (\kappa +\lambda )+6}}\right)$ & $\frac{4 \lambda -2 \kappa }{\sqrt{3}
   \sqrt{\kappa  (\kappa +\lambda )+6}}$ & $-\frac{\kappa +4 \lambda }{\sqrt{3} \sqrt{\kappa 
   (\kappa +\lambda )+6}}$ &$   \delta_+ $& $ \delta_-$ & saddle \\
$\bar{G}$ & $-\frac{\pi}{2}$ & $\frac{4 \lambda -2 \kappa }{\sqrt{3} \sqrt{\kappa  (\kappa +\lambda )+6}}$ & $0$ & $\delta_+$ & $\delta_-$  & saddle\\   
$H$  & $0$ & $\frac{2 (\kappa -2 \lambda )}{\sqrt{3}
   \sqrt{\kappa  (\kappa +\lambda )+6}}$ & $-\frac{\kappa +4 \lambda }{\sqrt{3} \sqrt{\kappa 
   (\kappa +\lambda )+6}}$ &  $\Delta_+ $& $\Delta_-$& saddle \\ 
$H^*$ & $-\tan ^{-1}\left(\frac{\kappa +4 \lambda }{\sqrt{3} \sqrt{\kappa  (\kappa +\lambda )+6}}\right)$ & $  \frac{2 (\kappa -2 \lambda )}{\sqrt{3}
   \sqrt{\kappa  (\kappa +\lambda )+6}}$ & $\frac{\kappa +4 \lambda }{\sqrt{3} \sqrt{\kappa 
   (\kappa +\lambda )+6}}$ &  $ \Delta_+ $& $  \Delta_-$& saddle \\ 
   $\bar{H}$ & $-\frac{\pi}{2}$ & $ \frac{2(\kappa -2 \lambda )}{\sqrt{3} \sqrt{\kappa  (\kappa +\lambda )+6}}$ & $0$ &  $\Delta_+$ & $\Delta_- $ & saddle  \\\hline
    \end{tabular}
    \caption{Coordinates and eigenvalues and of the equilibrium points of system \eqref{final-system1}, \eqref{final-system2},\eqref{final-system3}
and \eqref{final-system4}. The quantities $\delta_\pm$ and $\Delta_\pm$  are defined by \eqref{delta} and \eqref{Lambda} respectively.}
    \label{tab:perts}
\end{table*}
\begin{align*}
G,H &: \left(\mp \frac{\sqrt{6}}{\sqrt{\kappa^2 + \kappa \lambda +6}}, \pm \frac{\kappa + \lambda}{\sqrt{\kappa^2 + \kappa \lambda +6}}, 0, 0 \right), \\
G^*, H^*  &: \Bigg(\mp\frac{\sqrt{6}}{\sqrt{\kappa^2 + \kappa \lambda +6}},  \pm \frac{\kappa + \lambda}{\sqrt{\kappa^2 + \kappa \lambda +6}}, 0, \pm \tan ^{-1}\left(\Gamma\right) \Bigg),
\end{align*}
 where $\Gamma=\left(\kappa +4 \lambda\right)/\left({\sqrt{3} \sqrt{\kappa  (\kappa +\lambda )+6}}\right)$.
 
\noindent The dynamics of the angular variable on the limit $\bar{Z}=1$ is governed by the equation $\theta  ^{\prime} = \cos ^2(\theta  )$. The stability is analyzed passing to the time $\tau$. 
The equilibrium points in this invariant set satisfy $\theta= -\frac{\pi}{2}+n\pi, \; n\in \mathbb{Z}$ and  they are (modulo $n\pi, \; n\in \mathbb{Z}$): 
\begin{align*}
\bar{A}&: \left(1, 1, 1, -\frac{\pi}{2}\right), \qquad \bar{B}: \left(-1, 1, 1, -\frac{\pi}{2}\right), \\
\bar{C}&: \left(1, -1, 1, -\frac{\pi}{2}\right), \qquad\bar{D}: \left(-1, -1, 1, -\frac{\pi}{2}\right), \\
\bar{O}_{\pm}&: \left(\pm1, 0, 1, -\frac{\pi}{2}\right), \qquad \bar{E}, \bar{F}: \left(\mp\frac{\lambda}{\sqrt{6}}, \pm 1, 1, -\frac{\pi}{2} \right), 
\\
\bar{G}, \bar{H}  &: \left(\mp \frac{\sqrt{6}}{\sqrt{\kappa^2 + \kappa \lambda +6}}, \pm \frac{\kappa + \lambda}{\sqrt{\kappa^2 + \kappa \lambda +6}},1, -\frac{\pi}{2} \right).
\end{align*}

\noindent  
The sources of the system are 
\begin{itemize}
\item $ A$ for $\kappa<0, \lambda>-\sqrt{6}$, 
\item $B$ for $\kappa>0, \lambda<\sqrt{6}$, 
\item $ \bar{C}$ for $\kappa <0, \lambda >\sqrt{6}$, 
\item $ \bar{D}$  for $\kappa >0, \lambda <-\sqrt{6}$, 
\item $O_{+}$ for $\kappa>0, \lambda>-\kappa$,  
\item $O_{-}$ for $\kappa<0, \lambda<-\kappa$, 
\item $F^*$  for $\abs{\lambda}<\sqrt{2},  \kappa \lambda +\lambda^2 -6 <0$, and 
\item $\bar{F}$ for $\sqrt{2}<\abs{\lambda}<\sqrt{6},   \kappa \lambda +\lambda^2 -6 <0$.  
\end{itemize}
\noindent 
The sinks are 
\begin{itemize}
\item $ \bar{A}$ for $\kappa >0, \lambda <-\sqrt{6}$, 
\item $\bar{B}$  for $\kappa <0, \lambda >\sqrt{6}$, 
\item $C$ for $\kappa>0, \lambda<\sqrt{6}$, 
\item $D$ for $\kappa<0, \lambda>-\sqrt{6}$, 
\item $O_{+}$ for $\kappa<0, \lambda<-\kappa$, 
\item $O_{-}$ for $\kappa>0, \lambda>-\kappa$,  
\item $E^*$  for $\abs{\lambda}<\sqrt{2},  \kappa \lambda +\lambda^2 -6 <0$, and 
\item $\bar{E}$ for $\sqrt{2}<\abs{\lambda}<\sqrt{6},   \kappa \lambda +\lambda^2 -6 <0$.
\end{itemize}

\begin{table}[]
    \centering
    \begin{tabular}{|c|c|c|c|}
    \hline
        $\Phi$ & $ h$ & $\bar{Z}$ & $ \theta$    \\ \hline
       $-0.25$ &$-1$ & $0.25$ &$ \pi/4$\\
      $0.55$ &$ -1$ &$ 0.35$ &$ 3\pi/4$\\
      $-0.85$ &$ -1$ &$ 0.45$ &$ 7\pi/4$\\
      $0.25$ &$ 0$ &$ 0.55$ &$ \pi/4$\\
      $0.5$ & $0$ & $ 0$ & $\pi/4$\\
      $-0.5$ & $0$ & $0$ & $\pi/4$ \\
      $-0.55$ & $0$ & $0.65$ & $3\pi/4$\\
      $0.05$ & $ 0$ & $ 0$ & $ 3\pi/4$\\
      $0.85$ & $0$ & $0.75$ & $7\pi/4$\\
      $-0.25$ & $ 1$ & $ 0.55$ & $\pi/4$\\
      $0.55$ & $1$ & $0.65$ & $3\pi/4$\\
      $-0.85$ & $ 1$ & $ 0.75$ & $7\pi/4$\\
      $-0.05$ & $1$ & $ 0.95$ & $ 7\pi/4$\\\hline
    \end{tabular}
    \caption{Initial conditions for the third plot \ref{Plot_3}. }
    \label{ini}
\end{table}

\begin{figure*}[t!]
	\centering
	\subfigure[Projections $(\cos \theta, \sin \theta, \bar{Z})$. Invariant set $h=1$. We use the point's label $I: (\bar{Z}, \theta)=\left(1,-\frac{\pi}{2}+n\pi\right),\; n\in \mathbb{Z}$.]{\includegraphics[scale=0.28]{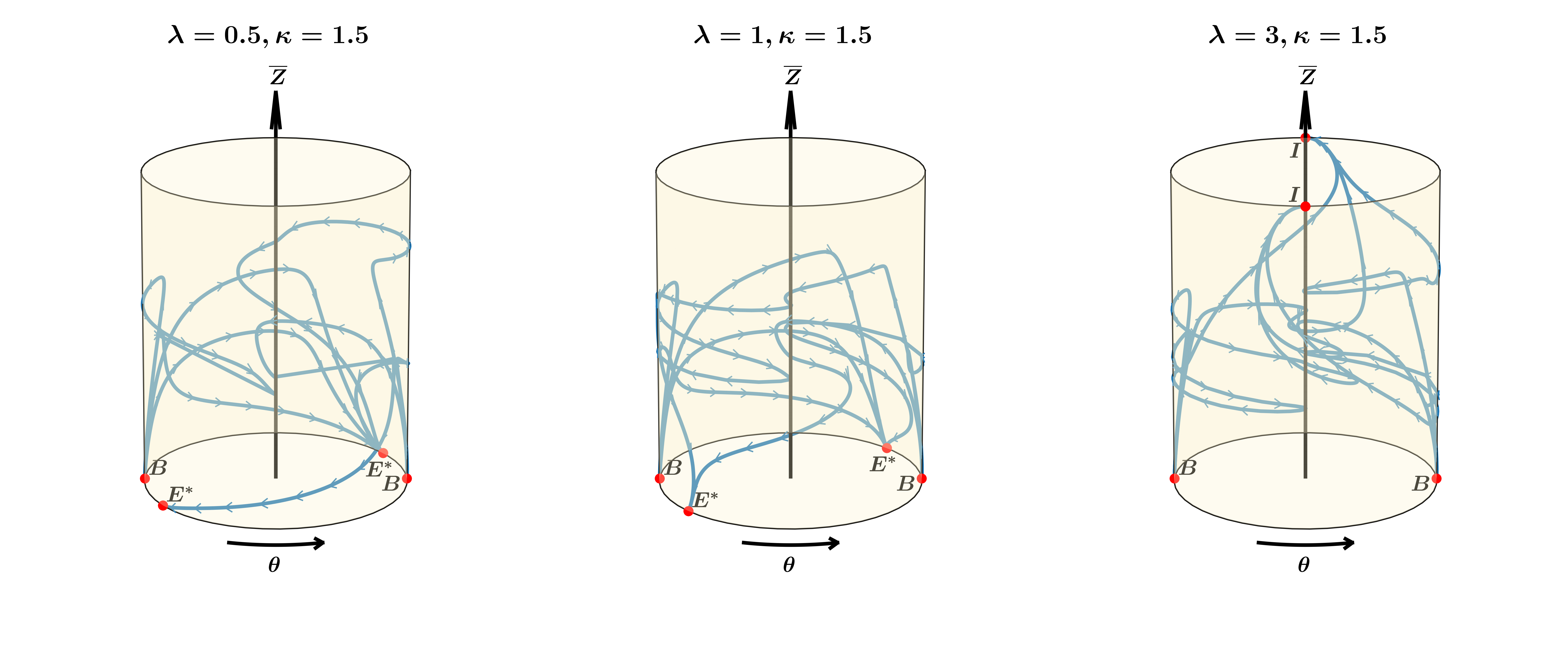}}
	\subfigure[Projections $(\cos \theta, \sin \theta, \bar{Z})$. Invariant set $h=-1$. We use the point's label $I: (\bar{Z}, \theta)=\left(1,-\frac{\pi}{2}+n\pi\right),\; n\in \mathbb{Z}$.]{\includegraphics[scale=0.28]{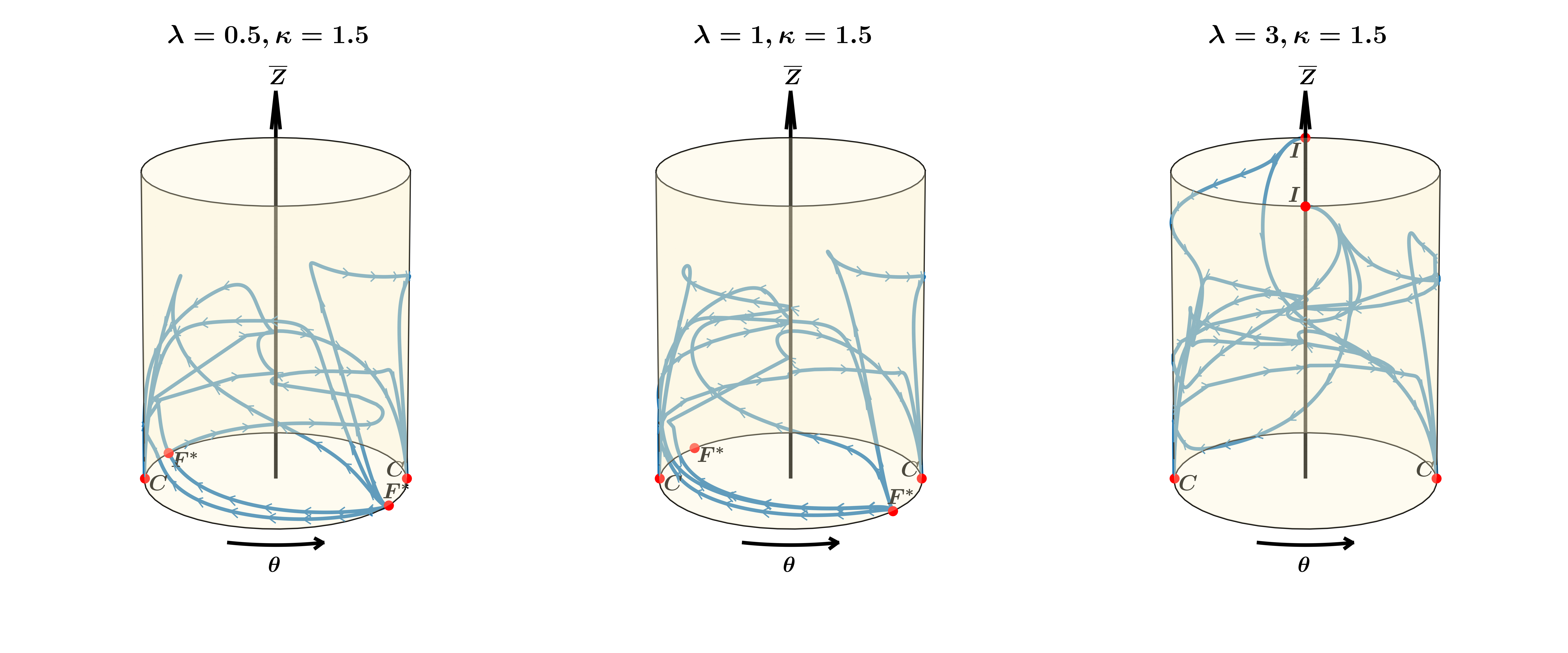}}
    \subfigure[\label{Plot_3} Projections $(h, \Phi, \bar{Z})$. We consider the initial conditions of Tab. \ref{ini} ]{\includegraphics[scale=0.28]{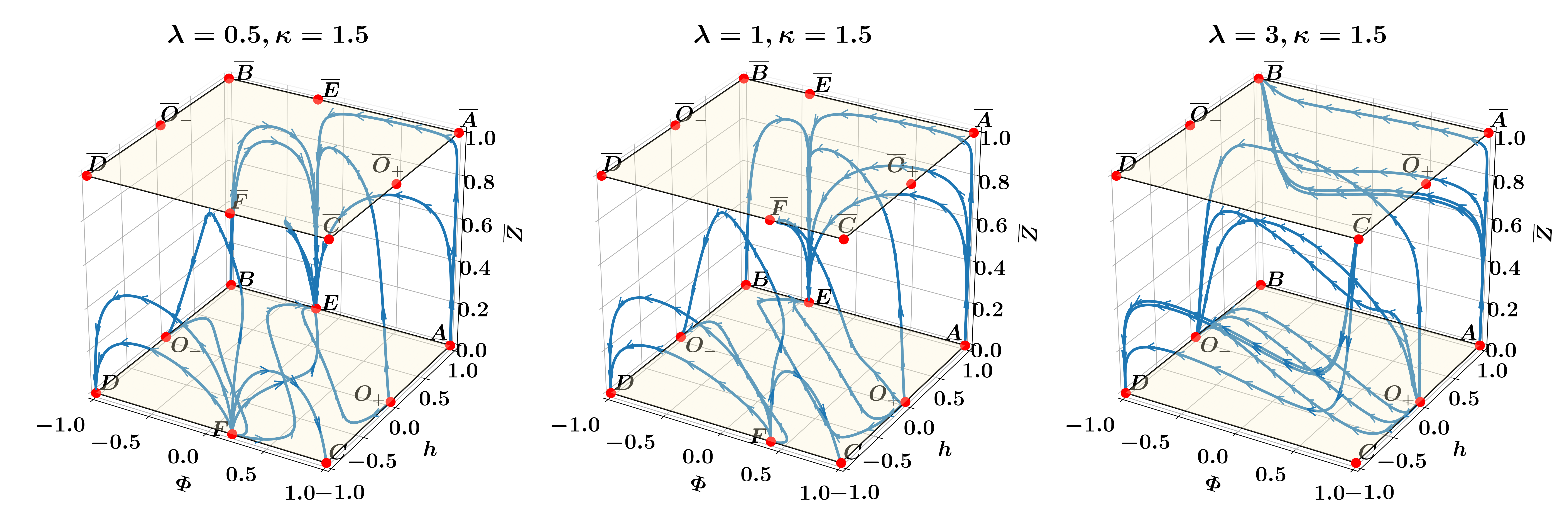}}
    \caption{ \label{fig:Proj} Projections  of solutions of  the system \eqref{final-system1}, \eqref{final-system2},\eqref{final-system3} and \eqref{final-system4}. }
\end{figure*}

In Figs. \ref{fig:Proj} we present projections of solutions of the system \eqref{final-system1}, \eqref{final-system2},\eqref{final-system3} and \eqref{final-system4} for different initial conditions, which display the behaviour of equilibrium points of interest. For the third plot \ref{Plot_3}, we consider the initial conditions in Tab. \ref{ini}. 

The following results are valid for any equilibrium points $P$, $P^*$ and $\bar{P}$.

Evaluating \eqref{evolve_v}, \eqref{evolve_omega}, \eqref{evolve_Sigma2} and \eqref{evolve_Xi} at the equilibrium points, and integrating, we have 
the asymptotic expansions
\begin{align*}
& A: \left\{\begin{array}{cc}
    \delta\phi  =\frac{1}{4} \left(\sqrt{6} {\Xi_0}+e^{-\frac{4 \tau }{\sqrt{3}}} \left(3 {\varphi_0}-\sqrt{6} {\Xi_0}\right)+{\varphi_0}\right)\\
\delta\psi e^{\kappa \phi} \frac{d{\psi}}{d\tau}  =\frac{1}{4} \left(e^{-\frac{4 \tau }{\sqrt{3}}} \left(\sqrt{6} {\Sigma_0}+3   {\omega_0}\right)-\sqrt{6} {\Sigma_0}+{\omega_0}\right)\\
   \mathcal{S}_\phi  = \frac{1}{8} \left(6 \Xi_{0}+e^{-\frac{4 \tau }{\sqrt{3}}} \left(2 {\Xi_0}-\sqrt{6} {\varphi_0}\right)+\sqrt{6}  {\varphi_0}\right)\\
   \mathcal{S}_\psi  =\frac{1}{8} \left(e^{-\frac{4 \tau }{\sqrt{3}}} \left(2 {\Sigma_0}+\sqrt{6} {\omega_0}\right)+6 {\Sigma_0}-\sqrt{6} {\omega_0}\right)\end{array}\right.,
\\\\
& B: \left\{\begin{array}{cc}
   \delta\phi  =\frac{1}{4} \left(-\sqrt{6} {\Xi_0}+e^{-\frac{4 \tau }{\sqrt{3}}} \left(\sqrt{6} {\Xi_0}+3    {\varphi_0}\right)+{\varphi_0}\right)\\
\delta\psi e^{\kappa \phi} \frac{d{\psi}}{d\tau}  =\frac{1}{4} \left(e^{-\frac{4 \tau }{\sqrt{3}}} \left(\sqrt{6} {\Sigma_0}+3 {\omega_0}\right)-\sqrt{6} {\Sigma_0}+{\omega_0}\right)\\
   \mathcal{S}_\phi  =\frac{1}{8} \left(6 {\Xi_0}+e^{-\frac{4 \tau }{\sqrt{3}}} \left(2 {\Xi_0}+\sqrt{6} {\varphi_0}\right)-\sqrt{6} {\varphi_0}\right)\\
   \mathcal{S}_\psi  =\frac{1}{8} \left(e^{-\frac{4 \tau }{\sqrt{3}}} \left(2 {\Sigma_0}+\sqrt{6} {\omega_0}\right)+6 {\Sigma_0}-\sqrt{6} {\omega_0}\right)
\end{array}\right., \\\\
& C : \left\{\begin{array}{cc}
    \delta\phi  =\frac{1}{4} \left(-\sqrt{6} {\Xi_0}+e^{\frac{4 \tau }{\sqrt{3}}} \left(\sqrt{6} {\Xi_0}+3   {\varphi_0}\right)+{\varphi_0}\right)\\
\delta\psi e^{\kappa \phi} \frac{d{\psi}}{d\tau}  =\frac{1}{4} \left(e^{\frac{4 \tau }{\sqrt{3}}} \left(3 {\omega_0}-\sqrt{6} {\Sigma_0}\right)+\sqrt{6} {\Sigma_0}+{\omega_0}\right)\\
   \mathcal{S}_\phi  =\frac{1}{8} \left(6   {\Xi_0}+e^{\frac{4 \tau }{\sqrt{3}}} \left(2 {\Xi_0}+\sqrt{6} {\varphi_0}\right)-\sqrt{6}   {\varphi_0}\right)\\
   \mathcal{S}_\psi  =\frac{1}{8} \left(e^{\frac{4 \tau }{\sqrt{3}}} \left(2 {\Sigma_0}-\sqrt{6} {\omega_0}\right)+6 {\Sigma_0}+\sqrt{6} {\omega_0}\right)
\end{array}\right.,\\\\
&  D: \left\{\begin{array}{cc}
   \delta\phi  =\frac{1}{4} \left(\sqrt{6} {\Xi_0}+e^{\frac{4 \tau }{\sqrt{3}}} \left(3 {\varphi_0}-\sqrt{6} {\Xi_0}\right)+{\varphi_0}\right)\\
\delta\psi e^{\kappa \phi} \frac{d{\psi}}{d\tau}  =\frac{1}{4} \left(e^{\frac{4 \tau }{\sqrt{3}}} \left(3 {\omega_0}-\sqrt{6}   {\Sigma_0}\right)+\sqrt{6} {\Sigma_0}+{\omega_0}\right)\\
   \mathcal{S}_\phi  =\frac{1}{8} \left(6 {\Xi_0}+e^{\frac{4 \tau }{\sqrt{3}}} \left(2 {\Xi_0}-\sqrt{6} {\varphi_0}\right)+\sqrt{6}   {\varphi_0}\right)\\
   \mathcal{S}_\psi  = \frac{1}{8} \left(e^{\frac{4 \tau }{\sqrt{3}}} \left(2 {\Sigma_0}-\sqrt{6} {\omega_0}\right)+6 {\Sigma_0}+\sqrt{6} {\omega_0}\right)
\end{array}\right.,
\\\\
& O_+ : \left\{\begin{array}{cc}
  \delta\phi  = {\varphi_0} e^{-\frac{\kappa  \tau }{\sqrt{2}}}\\
\delta\psi e^{\kappa \phi} \frac{d{\psi}}{d\tau}  =e^{-\frac{\kappa  \tau }{\sqrt{2}}} \left(e^{\sqrt{2} \kappa    \tau } ({\omega_0}-{\varphi_0})+{\varphi_0}\right)\\
   \mathcal{S}_\phi  = \frac{\kappa  {\Xi_0}+{\varphi_0}   \left(1-e^{-\frac{\kappa  \tau }{\sqrt{2}}}\right)}{\kappa }\\
   \mathcal{S}_\psi  = \frac{\kappa  {\Sigma_0}-{\omega_0} e^{\frac{\kappa  \tau }{\sqrt{2}}}+2 {\varphi_0} \cosh \left(\frac{\kappa  \tau   }{\sqrt{2}}\right)-2 {\varphi_0}+{\omega_0}}{\kappa }
\end{array}\right.,\\\\
& O_-: \left\{\begin{array}{cc}
  \delta\phi ={\varphi_0} e^{\frac{\kappa  \tau }{\sqrt{2}}}\\
\delta\psi e^{\kappa \phi} \frac{d{\psi}}{d\tau}  =e^{-\frac{\kappa  \tau }{\sqrt{2}}} \left({\varphi_0}   \left(1-e^{\sqrt{2} \kappa  \tau }\right)+{\omega_0}\right)\\
   \mathcal{S}_\phi  =\frac{\kappa  {\Xi_0}+{\varphi_0}   \left(1-e^{\frac{\kappa  \tau }{\sqrt{2}}}\right)}{\kappa }\\
   \mathcal{S}_\psi  = \frac{\kappa  {\Sigma_0}+{\omega_0} \left(e^{-\frac{\kappa  \tau }{\sqrt{2}}}-1\right)+2 {\varphi_0} \cosh \left(\frac{\kappa  \tau   }{\sqrt{2}}\right)-2 {\varphi_0}}{\kappa }
\end{array}\right.,
\end{align*}

\newpage 
\begin{strip}
\begin{alignat*}{1}
& E: \left\{\begin{array}{cc}
   \delta\phi  =\frac{\lambda  e^{-\frac{\left(\lambda ^2+2\right) \tau }{2 \sqrt{3}}} (2 {\Xi_0}+\lambda  {\varphi_0})+2   ({\varphi_0}-\lambda  {\Xi_0})}{\lambda ^2+2}\\
\delta\psi e^{\kappa \phi} \frac{d{\psi}}{d\tau}  =\frac{e^{-\frac{\left(-\lambda ^2+\sqrt{\lambda ^4-12 \lambda    ^2+100}+14\right) \tau }{4 \sqrt{3}}} \left(3 {\omega_0} \left(\left(\sqrt{\lambda ^4-12 \lambda
   ^2+100}-10\right) e^{\frac{\sqrt{\lambda ^4-12 \lambda ^2+100} \tau }{2 \sqrt{3}}}+\sqrt{\lambda ^4-12 \lambda   ^2+100}+10\right)-\lambda ^2 \left(e^{\frac{\sqrt{\lambda ^4-12 \lambda ^2+100} \tau }{2 \sqrt{3}}}-1\right)
   \left(2 \sqrt{6} {\Sigma_0}-3 {\omega_0}\right)\right)}{6 \sqrt{\lambda ^4-12 \lambda   ^2+100}}\\
   \mathcal{S}_\phi  =\frac{e^{-\frac{\left(\lambda ^2+2\right) \tau }{2 \sqrt{3}}} (2 {\Xi_0}+\lambda    {\varphi_0})+\lambda  (\lambda  {\Xi_0}-{\varphi_0})}{\lambda ^2+2}\\
   \mathcal{S}_\psi  = \frac{e^{-\frac{\left(-\lambda   ^2+\sqrt{\lambda ^4-12 \lambda ^2+100}+14\right) \tau }{4 \sqrt{3}}} \left({\Sigma_0} \left(\lambda
   ^2+\left(-\lambda ^2+\sqrt{\lambda ^4-12 \lambda ^2+100}+10\right) e^{\frac{\sqrt{\lambda ^4-12 \lambda ^2+100}   \tau }{2 \sqrt{3}}}+\sqrt{\lambda ^4-12 \lambda ^2+100}-10\right)-2 \sqrt{6} {\omega_0}
   \left(e^{\frac{\sqrt{\lambda ^4-12 \lambda ^2+100} \tau }{2 \sqrt{3}}}-1\right)\right)}{2 \sqrt{\lambda ^4-12   \lambda ^2+100}}
\end{array}\right.,
\\\\
& F: \left\{\begin{array}{cc}
 \delta\phi  =\frac{\lambda  e^{\frac{\left(\lambda ^2+2\right) \tau }{2 \sqrt{3}}} (2 {\Xi_0}+\lambda  {\varphi_0})+2
   ({\varphi_0}-\lambda  {\Xi_0})}{\lambda ^2+2},\\
\delta\psi e^{\kappa \phi} \frac{d{\psi}}{d\tau}  =\frac{e^{-\frac{\left(\lambda ^2+\sqrt{\lambda ^4-12 \lambda
   ^2+100}-14\right) \tau }{4 \sqrt{3}}} \left(3 {\omega_0} \left(\left(\sqrt{\lambda ^4-12 \lambda
   ^2+100}+10\right) e^{\frac{\sqrt{\lambda ^4-12 \lambda ^2+100} \tau }{2 \sqrt{3}}}+\sqrt{\lambda ^4-12 \lambda
   ^2+100}-10\right)-\lambda ^2 \left(e^{\frac{\sqrt{\lambda ^4-12 \lambda ^2+100} \tau }{2 \sqrt{3}}}-1\right)
   \left(2 \sqrt{6} {\Sigma_0}+3 {\omega_0}\right)\right)}{6 \sqrt{\lambda ^4-12 \lambda  ^2+100}},\\
   \mathcal{S}_\phi  = \frac{e^{\frac{\left(\lambda ^2+2\right) \tau }{2 \sqrt{3}}} (2 {\Xi_0}+\lambda 
   {\varphi_0})+\lambda  (\lambda  {\Xi_0}-{\varphi_0})}{\lambda ^2+2} ,\\
   \mathcal{S}_\psi  =\frac{e^{-\frac{\left(\lambda
   ^2+\sqrt{\lambda ^4-12 \lambda ^2+100}-14\right) \tau }{4 \sqrt{3}}} \left({\Sigma_0} \left(-\lambda
   ^2+\left(\lambda ^2+\sqrt{\lambda ^4-12 \lambda ^2+100}-10\right) e^{\frac{\sqrt{\lambda ^4-12 \lambda ^2+100}
   \tau }{2 \sqrt{3}}}+\sqrt{\lambda ^4-12 \lambda ^2+100}+10\right)-2 \sqrt{6} {\omega_0}
   \left(e^{\frac{\sqrt{\lambda ^4-12 \lambda ^2+100} \tau }{2 \sqrt{3}}}-1\right)\right)}{2 \sqrt{\lambda ^4-12
   \lambda ^2+100}}.
\end{array}\right.
\end{alignat*}
\end{strip}
For brevity, we have omitted the expansions for points $G$ and $H$.

\section{Discussion}

The most important physical consequence of the expanding quintom models is that in the parameter range $\kappa>0,0<\lambda<\sqrt{2},\kappa\lambda+\lambda^2-6<0$, the orbits allow for two inflationary periods. The initial one is when approaching the saddle $H$, which is inflationary, with both fields contributing. Then the orbits evolve to the sink $E$, corresponding to single-field exponential potential inflation. In curved cosmologies, the points' local stability and inflationary property that allow for two periods of inflation are unchanged. Namely, in curved models, the saddle is inflationary, and the sink is inflationary, corresponding to the equilibrium points $H$ and $E$, respectively. In addition, we describe a set of bouncing solutions of the model and the evolution of cosmological linear perturbations is examined in detail. 
In particular, we have investigated an autonomous system of nonlinear first-order ordinary differential equations, where the state space $S$ has a product structure $S = B \times P$. Here $B$ is the background state space, which describes the dynamics of an FLRW background, and $P$ is the perturbation state space, containing gauge invariant variables that describe linear cosmological perturbations. 

We have identified the sources $ A$ for $\kappa<0, \lambda>-\sqrt{6}$, $B$ for $\kappa>0, \lambda<\sqrt{6}$, $ \bar{C}$
for $\kappa <0, \lambda >\sqrt{6}$, $ \bar{D}$ 
for $\kappa >0, \lambda <-\sqrt{6}$, $O_{+}$ for $\kappa>0, \lambda>-\kappa$ and $O_{-}$
for $\kappa<0, \lambda<-\kappa$,  $F^*$  for $\abs{\lambda}<\sqrt{2},  \kappa \lambda +\lambda^2 -6 <0$, and $\bar{F}$
 for $\sqrt{2}<\abs{\lambda}<\sqrt{6},   \kappa \lambda +\lambda^2 -6 <0$.  The sinks, which are  $ \bar{A}$ for $\kappa >0, \lambda <-\sqrt{6}$, $\bar{B}$
 for $\kappa <0, \lambda >\sqrt{6}$, $C$ for $\kappa>0, \lambda<\sqrt{6}$, $D$ for $\kappa<0, \lambda>-\sqrt{6}$, $O_{+}$
for $\kappa<0, \lambda<-\kappa$, $O_{-}$ for $\kappa>0, \lambda>-\kappa$,   $E^*$  for $\abs{\lambda}<\sqrt{2},  \kappa \lambda +\lambda^2 -6 <0$, $\bar{E}$
 for $\sqrt{2}<\abs{\lambda}<\sqrt{6},   \kappa \lambda +\lambda^2 -6 <0$.

In future work, we will present a complete dynamical analysis of this model with spatial curvature, and for a second quintom model \cite{PaperII}.

\section*{Acknowledgments}
 A. C. was supported by NSERC of Canada, and   G. L.  was funded by Vicerrectoría de
Investigación y Desarrollo Tecnológico (Vridt) at Universidad Católica del Norte through
Concurso De Pasantías De Investigación Año 2022, Resolución Vridt N° 040/2022 and
through Resolución Vridt N° 054/2022. G.L. thanks the support of Núcleo de Investigación Geometría Diferencial y Aplicaciones, Resolución Vridt N°096/2022. We also thanks Esteban Gonz\'alez for preparing Figs. \ref{fig:Proj}.

\end{document}